\documentclass[journal]{IEEEtran}

\usepackage{cite}
\usepackage{amsmath,amssymb,amsfonts}
\usepackage{algorithmic}
\usepackage[linesnumbered,ruled,vlined]{algorithm2e}
\usepackage{graphicx,color}
\usepackage{makecell}
\usepackage{booktabs}
\usepackage{multicol}
\usepackage{CJKutf8}
\usepackage{multirow} 
\usepackage{cite,amsfonts,amssymb,color, mathtools,setspace,comment,float,array}
\usepackage{textcomp}
\usepackage[hidelinks, breaklinks = true]{hyperref} 
\usepackage{amsmath,amssymb,amsfonts}
\usepackage{algorithmic}
\usepackage{graphicx}
\usepackage{textcomp}
\usepackage{array}
\usepackage{makecell}
\usepackage{xcolor}
\usepackage{subcaption}
\def\BibTeX{{\rm B\kern-.05em{\sc i\kern-.025em b}\kern-.08em
    T\kern-.1667em\lower.7ex\hbox{E}\kern-.125emX}}
\AtBeginDocument{\definecolor{ojcolor}{cmyk}{0.93,0.59,0.15,0.02}}

\begin{document}
\begin{CJK}{UTF8}{gbsn}

\title{A Survey of New Mid-Band/FR3 for 6G: Channel Measurement, Characterization and Modeling in Outdoor Environment }

%
%
%



\author{Haiyang~Miao, Jianhua~Zhang, Pan~Tang, Jie~Meng, Qi~Zhen, Ximan~Liu, Enrui~Liu, Peijie~Liu, Lei Tian, Guangyi Liu

		
\thanks{This work was supported by National Key R\&D Program of China (2023YFB2904802), Natural Science Foundation of Beijing-Xiaomi Innovation Joint Foundation (L243002), National Natural Science Foundation of China (62101069 \& 62201086), and Beijing University of Posts and Telecommunications-China Mobile Research Institute Joint Innovation Center. (\itshape{Corresponding author: Jianhua Zhang.})}

\thanks{H. Miao, J. Zhang, P. Tang, and L. Tian are with State Key Laboratory of Networking and Switching Technology, Beijing University of Posts and Telecommunications, Beijing 100876, China (e-mail: hymiao@bupt.edu.cn; jhzhang@bupt.edu.cn; tangpan27@bupt.edu.cn; tianlbupt@bupt.edu.cn).}

}

\markboth{Journal of \LaTeX\ Class Files,~Vol.~14, No.~8, August~2021}%
{Shell \MakeLowercase{\textit{et al.}}: A Sample Article Using IEEEtran.cls for IEEE Journals}

%

\maketitle

\begin{abstract}
	
The new mid-band (6-24 GHz) has attracted significant attention from both academia and industry, which is the spectrum with continuous bandwidth that combines the coverage benefits of low frequency with the capacity advantages of high frequency. Since outdoor environments represent the primary application scenario for mobile communications, this paper presents the first comprehensive review and summary of multi-scenario and multi-frequency channel characteristics based on extensive outdoor new mid-band channel measurement data, including UMa, UMi, and O2I. Specifically, a survey of the progress of the channel characteristics is presented, such as path loss, delay spread, angular spread, channel sparsity, capacity and near-field spatial non-stationary characteristics. Then, considering that satellite communication will be an important component of future communication systems, we examine the impact of clutter loss in air-ground communications. Our analysis of the frequency dependence of mid-band clutter loss suggests that its impact is not significant. Additionally, given that penetration loss is frequency-dependent, we summarize its variation within the FR3 band. Based on experimental results, comparisons with the standard model reveal that while the 3GPP TR 38.901 model remains a useful reference for penetration loss in wood and glass, it shows significant deviations for concrete and glass, indicating the need for further refinement. In summary, the findings of this survey provide both empirical data and theoretical support for the deployment of mid-band in future communication systems, as well as guidance for optimizing mid-band base station deployment in the outdoor environment. This survey offers the reference for improving standard models and advancing channel modeling.

\end{abstract}


\IEEEpeerreviewmaketitle

\section{INTRODUCTION}

With each generation of mobile communication, new frequency bands are allocated to meet the growing demand for connectivity, higher data rates, and enhanced network capacity, ensuring that the system evolves to support modern communication needs. 
In June 2023, the 44th meeting of the International Telecommunication Union - Radiocommunication Sector (ITU-R) Working Party 5D (WP 5D) defined the overall objectives and key trends for sixth-generation (6G) mobile networks. At this meeting, six usage scenarios for International Mobile Telecommunications (IMT) for 2030 (IMT-2030) and beyond were proposed \cite{recommendation2023framework}, emphasizing the growing demand for ultra-high data rates and extensive coverage. However, meeting these demands is likely to require the allocation of additional spectrum resources to communication systems.

\begin{figure}[!htbp]
    \centering
    \includegraphics[width=0.45\textwidth]{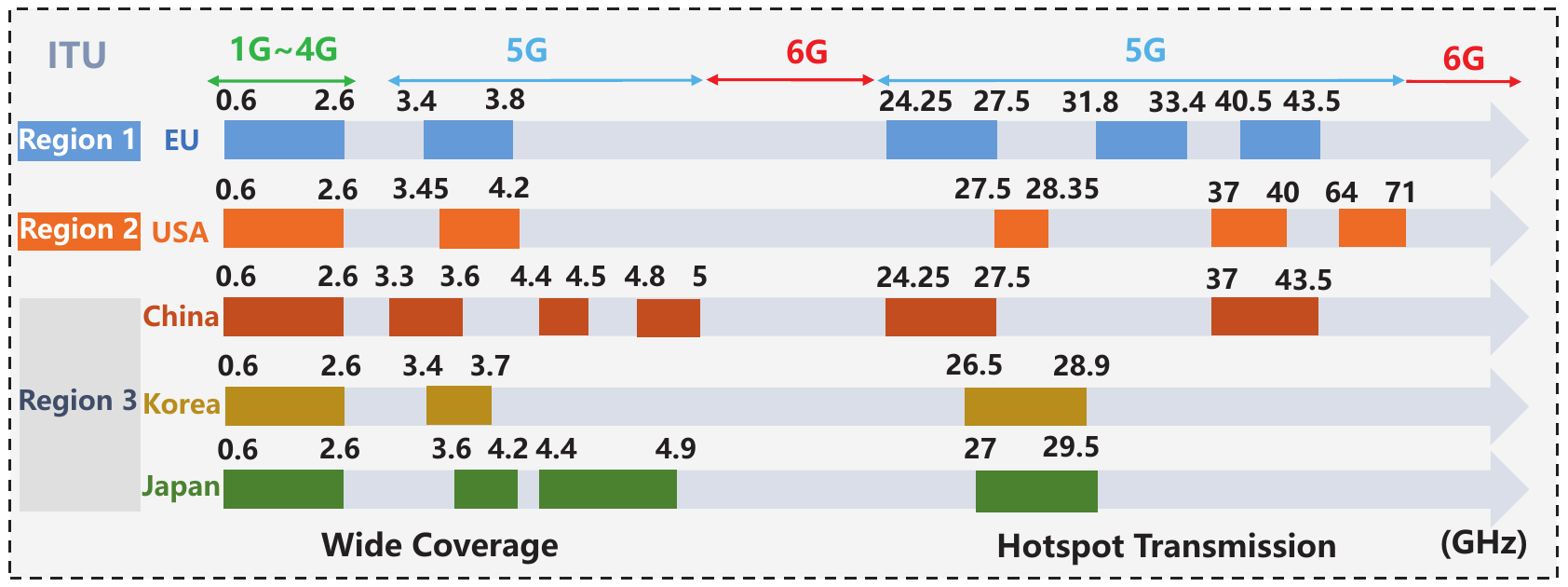}
    \caption{The operating and potential spectrum in ITU regions}
    \label{fig:figure1}
\end{figure}

The global frequency spectrum is managed by dividing the world into three ITU regions. Fig.~\ref{fig:figure1} illustrates the spectrum allocation for selected countries within these regions \cite{zhang2024new}. From the first generation (1G) to 5G, it is evident that mobile communication frequency bands are allocated within the sub-6 GHz range as well as above 24 GHz. In the forthcoming 6G era, collaborative communications across multiple frequency bands are expected to enable the efficient utilization of both low- and high-frequency resources, thereby offering significant application potential. However, as the frequency span increases, the variations in channel characteristics become more pronounced, presenting substantial challenges for multi-band wireless networking. Consequently, it is advisable to limit the frequency span and focus on adjacent frequency bands. These challenges highlight the critical importance of carefully selecting frequency bands for 6G development. 

The Sub-6 GHz frequency bands in fifth-generation (5G) networks can be reused through spectrum refarming; however, there is also a need to identify wider bandwidth spectrum resources. Currently, 6G development 
 mainly focuses on the new mid-band spectrum (6-24 GHz)\cite{zhang2024new}. In June 2022, the 3rd Generation Partnership Project (3GPP) Radio Access Network (RAN) Plenary approved  a standard modification proposal for the 5925–7125 MHz band, officially including it into 5G-Advanced Release 18. The 2023 World Radiocommunication Conference (WRC-23) designated the 6 GHz band for mobile use in all International Telecommunication Union (ITU) regions and set the agenda for the WRC-27. The new WRC cycle will prioritize the following frequency bands for International Mobile Telecommunications (IMT): 4400-4800 MHz, 7125–8400 MHz, and 14.8-15.35 GHz. In December 2023, the 3GPP Technical Specification Group (TSG) RAN Release 19 initiated discussions on research activities within the 7-24 GHz FR3 band. 6G is expected to utilize more abundant spectrum resources to achieve tightly integrated cross-band collaborative communication between low and high frequencies. Therefore, it is essential to study the frequency dependence of channels across such a wide frequency span, ranging from centimeter waves to millimeter waves.

The wireless channel serves as the transmission medium between the transmitter (Tx) and receiver (Rx). It plays a crucial role in determining the ultimate performance limit of mobile communication systems \cite{molisch2012wireless}. To support the innovative research and performance evaluation of key enabling technologies for 6G and to achieve the vision and goals of 6G, it is crucial to explore the novel channel characteristics of 6G in the mid-band spectrum and develop accurate channel models. Several channel measurements have been conducted to examine and investigate the new characteristics in the new mid-band spectrum.

\begin{table*}[!htbp]
\centering
\caption{Path loss, RMS DS, and RMS ASA model parameters for new mid-band in outdoor environment.}
\label{model parameters for mid-band}
\renewcommand{\arraystretch}{2}
	 \setlength{\tabcolsep}{1.25 mm}
\resizebox{\textwidth}{!}{
\begin{tabular}{|c|c|c|c|cc|cc|cc|c|}
\hline
\multirow{2}{*}{\textbf{S.N.}} & \multirow{2}{*}{\textbf{\begin{tabular}[c]{@{}c@{}}Freq.band\\ {[}GHz{]}\end{tabular}}} & \multirow{2}{*}{\textbf{\begin{tabular}[c]{@{}c@{}}Meas.Bw\\ {[}MHz{]}\end{tabular}}} & \multirow{2}{*}{\textbf{Scenario}}                           & \multicolumn{2}{c|}{\textbf{\begin{tabular}[c]{@{}c@{}}PLE\\ (CI,d0=1m)\end{tabular}}} & \multicolumn{2}{c|}{\textbf{RMS DS}}                                  & \multicolumn{2}{c|}{\textbf{RMS ASA}}                               & \multirow{2}{*}{Ref}      \\ \cline{5-10}
                               &                                                                                         &                                                                                       &                                                              & \multicolumn{1}{c|}{\textbf{LOS (n)}}                & \textbf{NLOS (n)}               & \multicolumn{1}{c|}{\textbf{LOS μ{[}ns{]}}} & \textbf{NLOS μ{[}ns{]}} & \multicolumn{1}{c|}{\textbf{LOS μ{[}°{]}}} & \textbf{NLOS μ{[}°{]}} &                           \\ \hline
\multirow{2}{*}{1}             & 6.5                                                                                     & 200                                                                                   & UMi                                                          & \multicolumn{1}{c|}{2.07}                            & 2.38                            & \multicolumn{1}{c|}{-}                      & 40.74                   & \multicolumn{1}{c|}{-}                     & -                      & \multirow{2}{*}{{[}4{]}}  \\ \cline{2-10}
                               & 15                                                                                      & 1000                                                                                  & UMi                                                          & \multicolumn{1}{c|}{2.01}                            & 2.59                            & \multicolumn{1}{c|}{-}                      & 21.38                   & \multicolumn{1}{c|}{-}                     & -                      &                           \\ \hline
2                              & 6                                                                                       & 200                                                                                   & UMa                                                          & \multicolumn{1}{c|}{-}                               & -                               & \multicolumn{1}{c|}{38.90}                  & 72.44                   & \multicolumn{1}{c|}{30.90}                 & 44.67                  & {[}9{]}                   \\ \hline
\multirow{2}{*}{3}             & \multirow{2}{*}{6}                                                                      & \multirow{2}{*}{200}                                                                  & UMi                                                          & \multicolumn{1}{c|}{-}                               & -                               & \multicolumn{1}{c|}{-}                      & -                       & \multicolumn{1}{c|}{36.31}                 & 54.95                  & \multirow{2}{*}{{[}12{]}} \\ \cline{4-10}
                               &                                                                                         &                                                                                       & UMa                                                          & \multicolumn{1}{c|}{-}                               & -                               & \multicolumn{1}{c|}{-}                      & -                       & \multicolumn{1}{c|}{45.71}                 & 83.18                  &                           \\ \hline
4                              & 6                                                                                       & 200                                                                                   & UMa                                                          & \multicolumn{1}{c|}{-}                               & -                               & \multicolumn{1}{c|}{30.90}                  & 79.43                   & \multicolumn{1}{c|}{-}                     & -                      & {[}13{]}                  \\ \hline
5                              & 6.75                                                                                    & 614                                                                                   & V2V                                                          & \multicolumn{1}{c|}{-}                               & -                               & \multicolumn{1}{c|}{19.7}                   & 49                      & \multicolumn{1}{c|}{28.5}                  & 42                     & {[}14{]}                  \\ \hline
6                              & 15                                                                                      & 4000                                                                                  & Outdoor                                                      & \multicolumn{1}{c|}{-}                               & -                               & \multicolumn{1}{c|}{24.55}                  & 93.3                    & \multicolumn{1}{c|}{9.33}                  & 40.74                  & {[}15{]}                  \\ \hline
\multirow{2}{*}{7}             & 6.75                                                                                    & 1000                                                                                  & UMi                                                          & \multicolumn{1}{c|}{1.89}                            & 3.25                            & \multicolumn{1}{c|}{29.1}                   & 35.6                    & \multicolumn{1}{c|}{19.05}                 & 31.62                  & \multirow{2}{*}{{[}16{]}} \\ \cline{2-10}
                               & 16.95                                                                                   & 1000                                                                                  & UMi                                                          & \multicolumn{1}{c|}{1.97}                            & 3.51                            & \multicolumn{1}{c|}{28.1}                   & 31.7                    & \multicolumn{1}{c|}{13.18}                 & 22.91                  &                           \\ \hline
\multirow{3}{*}{8}             & \multirow{3}{*}{18}                                                                     & \multirow{3}{*}{1000}                                                                 & \begin{tabular}[c]{@{}c@{}}UMi \\ Street Canyon\end{tabular} & \multicolumn{1}{c|}{-}                               & 3.1                             & \multicolumn{1}{c|}{-}                      & -                       & \multicolumn{1}{c|}{-}                     & -                      & \multirow{3}{*}{{[}17{]}} \\ \cline{4-10}
                               &                                                                                         &                                                                                       & \begin{tabular}[c]{@{}c@{}}UMi\\ Open Square\end{tabular}    & \multicolumn{1}{c|}{-}                               & 2.8                             & \multicolumn{1}{c|}{-}                      & -                       & \multicolumn{1}{c|}{-}                     & -                      &                           \\ \cline{4-10}
                               &                                                                                         &                                                                                       & UMa                                                          & \multicolumn{1}{c|}{-}                               & 2.9                             & \multicolumn{1}{c|}{-}                      & -                       & \multicolumn{1}{c|}{-}                     & -                      &                           \\ \hline
9                              & 10.1                                                                                    & 500                                                                                   & \begin{tabular}[c]{@{}c@{}}UMi\\ Street Canyon\end{tabular}  & \multicolumn{1}{c|}{2.0}                             & 3.0                             & \multicolumn{1}{c|}{19.95}                  & 38.90                   & \multicolumn{1}{c|}{29.51}                 & 58.88                  & {[}18{]}                  \\ \hline
\end{tabular}
}
\end{table*}

\begin{figure*}[ht]
    \centering
    \includegraphics[width=0.85\linewidth]{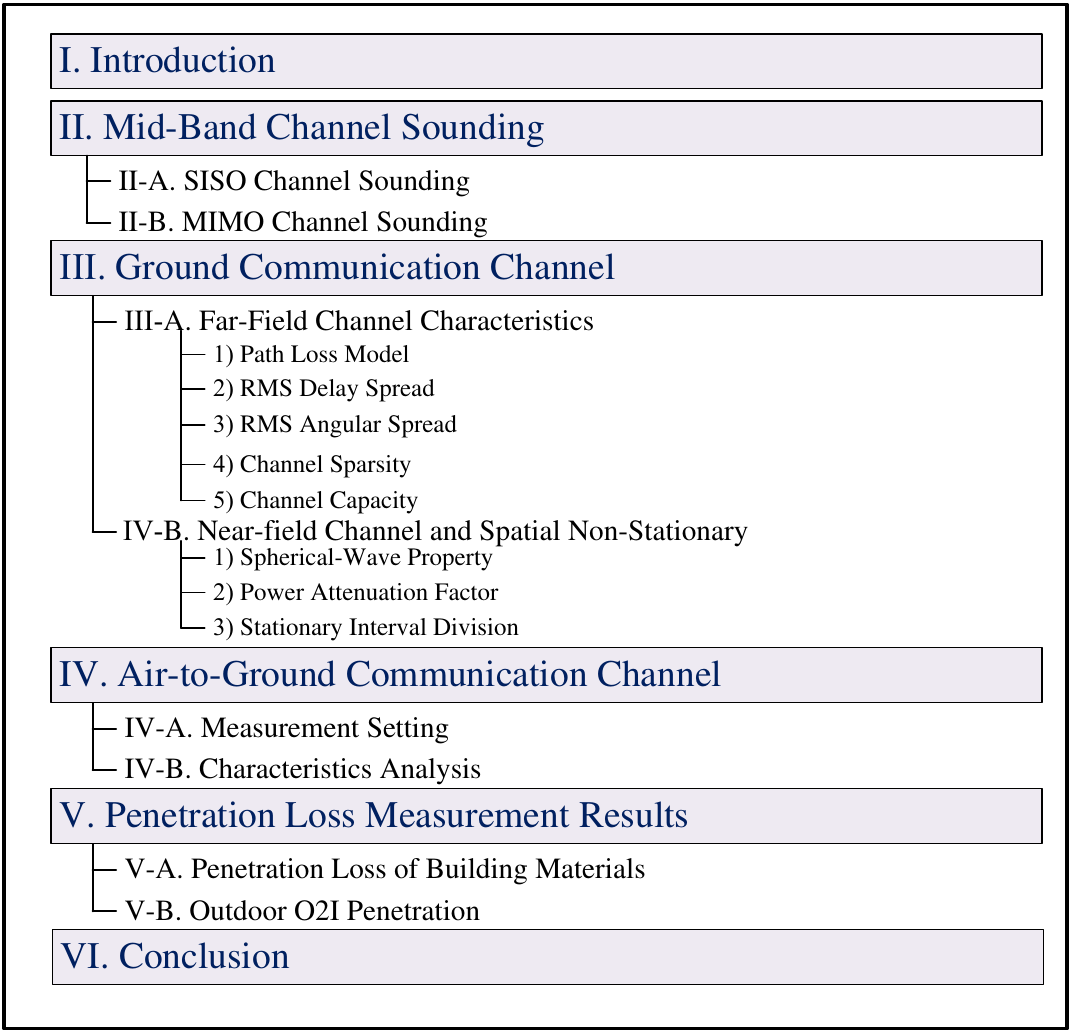}
    \caption{The organizational structure of this paper.}
    \label{fig:Structure}
\end{figure*}
Several channel measurement campaigns in outdoor environment have been summarized in the literature for the mid-band spectrum. Table \ref{model parameters for mid-band} presents the typical model parameters in the spatial, temporal, and angular domains obtained from these measurement campaigns. Measurements in \cite{miao2023Sub-6} were carried out at 3.3, 6.5, 15, and 28 GHz in a UMi scenario using a channel sounding platform based on the time-domain spread spectrum sliding correlation principle with a horn antenna as TX and an omnidirectional biconical antenna as RX. It was found that path loss increased with frequency, while delay spread tended to decrease. Moreover, NLoS path loss increased more rapidly with frequency, leading to the conclusion that the mid-band frequencies provided better signal coverage and stronger diffraction capabilities compared to 28 GHz. The \cite{Chen2024Experimental} conducted channel measurements at multiple mid-band frequencies (11.2–14.8 GHz) in a UMa scenario using the time-domain channel sounder based on a wide-band correlation , in which the TX opted for a horn antenna, while the RX employed antennas with three different half-power beam-widths (HPBW). The study found that, in LoS and OLoS scenarios, channel similarity across different frequencies was high, but as the HPBW difference or frequency difference increased, channel similarity decreased. Additionally, under large HPBW differences, increasing the signal frequency also led to reduced similarity. In \cite{Liu2023Empirical}, multi-frequency measurements were conducted at 3.3–7.5 GHz in a UMa scenario, revealing that clutter loss did not exhibit significant frequency dependency and varied noticeably between LoS and NLoS conditions. The \cite{liu2024experimentalanalysis} investigated penetration loss characteristics of materials such as wood, glass, foam, and concrete over the 4.5–15.5 GHz frequency range using a vector network analyzer (VNA)-based frequency-domain channel sounding system. It was found that penetration loss generally exhibited a linear relationship with frequency, except for glass, which displayed a nonlinear trend. Furthermore, the study highlighted certain limitations of the existing TR 38.901 model in predicting penetration loss within this frequency range \cite{3gpp38.901}. Measurements in \cite{Miao2024Measurement} were conducted in a UMa scenario over the 3–16 GHz frequency range using a channel sounding platform based on TDM-MIMO mode. At 6 GHz with a 200 MHz bandwidth in a LoS environment, the average root mean square (RMS) delay spread was 38.90 ns, and the average RMS angle spread was 30.90°. In an NLoS environment, the average RMS delay spread increased to 72.44 ns, and the average RMS angle spread rose to 44.67°. These results indicated that delay spread was larger in NLoS environments, consistent with the 3GPP TR 38.901 standard\cite{3gpp38.901}. The \cite{Wei2024Uma} performed XL-MIMO channel measurements at 13 GHz in a UMa scenario and analyzed near-field and spatial non-stationary characteristics. Significant multi-path angle shifts were observed at RX positions close to the transmitter, confirming the existence of spherical wavefronts. The distribution of multi-path power and angles in the array domain revealed spatial non-stationarity caused by partial blockages of the array by obstacles. The \cite{miao2024empiricalstudiespropagationcharacteristics} conducted XL-MIMO channel measurements at 6 GHz from near-field to far-field in an outdoor scenario. It was found that, regardless of whether the RX was in the near-field or transitioning to the far-field, path loss followed a logarithmic distribution with the RX-TX distance changing, but in the array domain, it exhibited a linear distribution. In the angular domain, the angular spread at the RX followed a linear trend with a small slope (absolute value less than 0.3) as the RX-TX distance increased. In contrast, the TX exhibited significant variations, including abrupt changes even with fluctuations exceeding 20°. In \cite{Zhang20183D}, field measurements were performed in three typical deployment scenarios, including O2I, UMi, and UMa at both 3.5 and 6 GHz frequencies with 200 MHz bandwidth. The effort reported Umi RMS angular spreads of 36.31° in LoS and 54.95° in NLoS, and UMa RMS angular spreads of 45.71°in LoS and 83.18°in NLoS. Measurements in \cite{Zheng2017Propagation} were performed in a UMa scenario at different frequencies (3.5 and 6 GHz) with different bandwidths (100 and 200 MHz). RMS delay spread values were illustrated as 30.90 ns in LoS and 79.43 ns in NLoS. The authors in \cite{Boban2019Multi} conducted Vehicle-to-vehicle (V2V) channel measurements in urban and highway scenarios at four different frequency bands (6.75, 30, 60, and 73 GHz). At 6.75 GHz with a 614 MHz bandwidth in a LoS environment, the average RMS delay spread was 19.7 ns, and the average RMS angle spread was 28.5°. In an NLoS environment, the average RS delay spread increased to 49 ns, and the average RMS angle spread rose to 42°. In \cite{Chen2017Measurement}, a measurement campaign for MIMO channel characterization was conducted at the center frequency of 15 GHz with a bandwidth of 4 GHz. Mean RMS delay spreads were reported as 24.55 ns in LoS and 9.33 ns in NLoS, and mean RMS angular spreads were 93.3°in LoS and 40.74°. The \cite{shakya2024urbanoutdoorpropagationmeasurements} presented an extensive UMi outdoor propagation measurement campaign at 6.75 GHz and 16.95 GHz conducted using the 1 GHz bandwidth sliding correlation channel sounder. The path loss exponents, RMS DS and RMS AS mean values were obtained. Authors in \cite{Sun2016Propagation} conducted channel propagation measurement campaigns from 2 GHz to 73.5 GHz in three typical outdoor scenarios, including UMi street canyon, UMi open square and UMa. The results showed that the one-parameter CI model had a very similar goodness of fit in both LoS and NLoS environments, as compared to the three-parameter ABG model. The measurements in \cite{Roivainen2017Parame} were performed in UMi scenario with a vector network analyzer at 10 GHz with 500 MHz bandwidth, and recorded the path loss exponents, RMS DS and RMS AS mean values in LOS and NLOS.

This paper presents a comprehensive study on radio propagation in outdoor environments, conducted at mid-band frequencies using a high-precision channel sounder.  Direct comparisons with 3GPP models are given in this paper, and offer valuable insights into the accuracy of models and how they might need to be revised for greater accuracy and broader applicability. The key contributions in this paper are as listed:

\begin{itemize}
    \item The wide-band channel sounder, specifically designed to operate in the mid-band (6–24 GHz range), is described in Section \ref{sec:II}. Additionally, detailed measurement methodology to capture the radio propagation behavior in the outdoor environment, and to conduct measurements is described in Section \ref{sec:II}.
\end{itemize}

\begin{itemize}
    \item Section \ref{sec:III} presents both the far-field and near-field channel characteristics in an outdoor environment. Firstly, path loss analysis  is conducted and resulting channel models are generated from sub-6GHz to mmWave bands (including 3.3, 6, 6.5, 13, 15, 28 GHz) using the close-in 1 m free space reference distance model. The results are compared with 3GPP models. Spatial-temporal statistics, including RMS DS and RMS ASA, extracted from the radio channel measurements at mid-band are also presented in Section \ref{sec:III} and compared with 3GPP models. Moreover, channel sparsity, channel capacity, and near-field channel characteristics are given in Section \ref{sec:III}.
\end{itemize}

\begin{itemize}
    \item Section \ref{sec:IV} : Clutter loss, a critical channel characteristic in air-ground communication, is discussed. In this paper, the channel measurement of clutter loss at 10-15 GHz is carried out. It is found that the frequency band has no obvious regularity to the clutter loss.
\end{itemize}

\begin{itemize}
    \item The results of penetration loss for building materials and outdoor O2I from FR1 to FR3 band is presented in Section \ref{sec:V}. Comparisons are made with 3GPP standard models for material penetration.
\end{itemize}

Fig. \ref{fig:Structure} illustrates the organization structure of this paper. Section \ref{sec:II} presents an overview of the SISO and MIMO channel sounder platform used for mid-frequency band channel measurements. Section \ref{sec:III} discusses mid-band outdoor communication scenarios including far-field and near-field propagation. Section \ref{sec:IV} provides an analysis of clutter loss in ground-to-air mid-band communication channels. Section \ref{sec:V} analyses the penetration loss within the mid-band frequency, covering the penetration loss of the material and the characteristics of the O2I penetration loss. Conclusions are given in Section \ref{sec:VI}.

\section{MID-BAND CHANNEL SOUNDING}
\label{sec:II}

\subsection{SISO Channel Sounding}

\par
\begin{figure}[htbp]
	\setlength{\abovecaptionskip}{0.3 cm}
	\centering
	\includegraphics[width=0.4\textwidth]{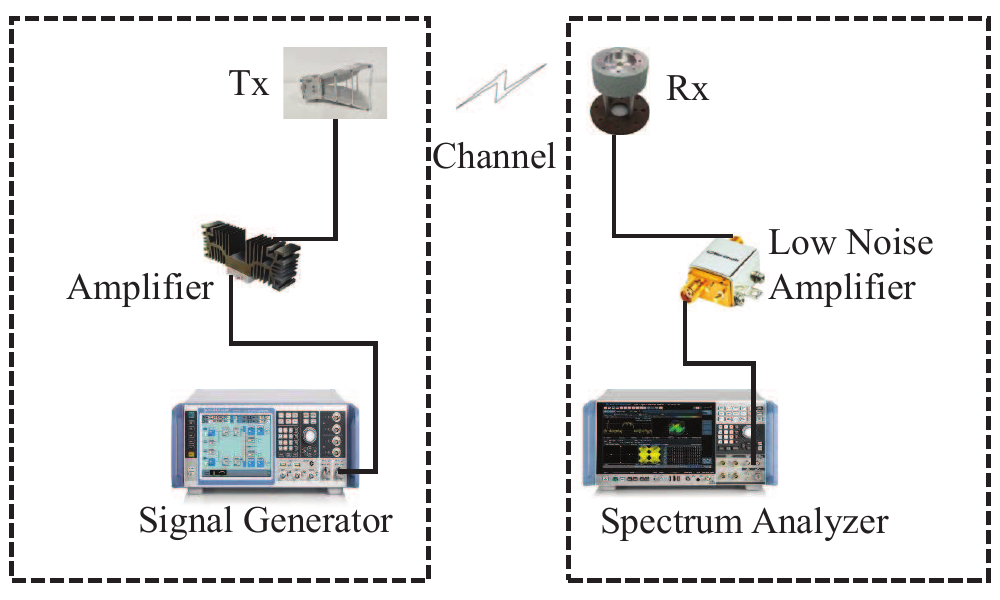}
\caption{The diagram of time-domain correlation channel measurement platform.}
	\label{fig:Figure1}
\end{figure}

 In Fig. \ref{fig:Figure1}, the single-input single-output (SISO) measurement platform operates on the principle of time-domain spread-spectrum sliding correlation \cite{miao2023Sub-6}. It uses a vector signal generator and a spectrum analyzer, thereby supporting wide-band channel measurements and capturing multi-path signals. To enhance the dynamic range, a 35-dB amplifier is installed at the Tx side, and a 20-dB low-noise amplifier is placed at the Rx side. This setup enables the system to measure path losses of at least 150 dB in each frequency band. To extract the pure channel impulse response, back-to-back calibration is performed to remove the effects of the measurement apparatus. Additionally, a frequency-switching program control is adopted to eliminate the influence of environmental changes and allow direct comparisons of instantaneous channel characteristics across different frequency bands.

\par The signal transmitter is realized using a Rohde \& Schwarz (R\&S) SMW 200A signal generator, which operates over a frequency range of 100 kHz to 40 GHz. Meanwhile, the Rohde \& Schwarz (R\&S) FSW43 spectrum analyzer is employed as the signal receiver and acquisition device, providing real-time display of frequency-domain signals and enabling real-time collection of time-domain IQ signals for subsequent analysis and processing.

\subsection{MIMO Channel Sounding}

\par
\begin{figure}[htbp]
	\xdef\xfigwd{\columnwidth}
	\setlength{\abovecaptionskip}{0.1 cm}
	\centering
	\begin{tabular}{c}
		\includegraphics[width=0.42\textwidth]{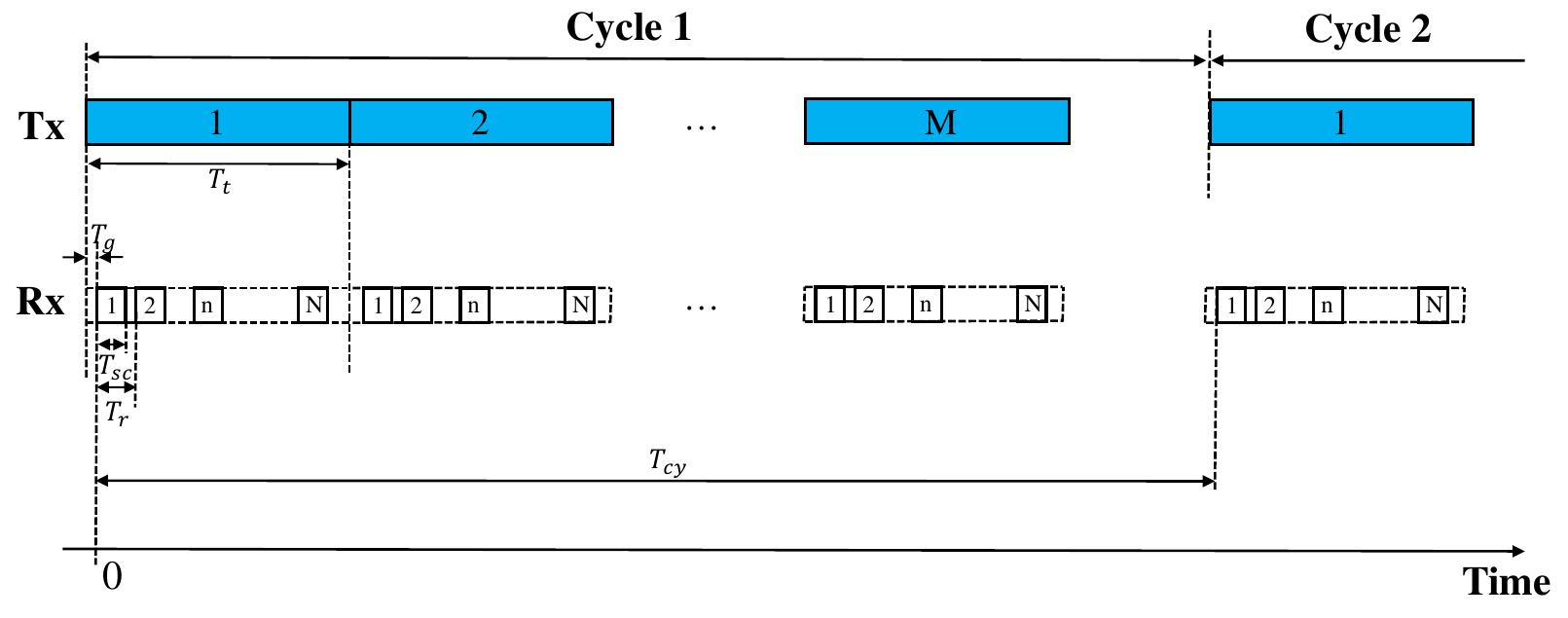}\\
		{\footnotesize\sf (a)} \\[3mm]
		\includegraphics[width=0.42\textwidth]{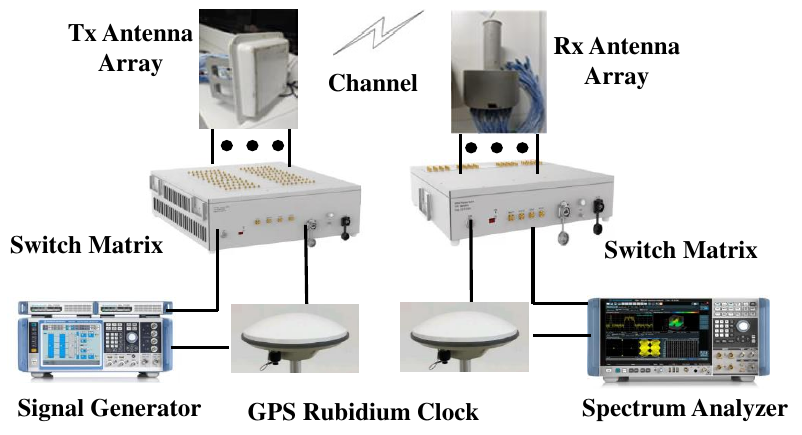}\\
		{\footnotesize\sf (b)} \\
	\end{tabular}
	\caption{ Massive MIMO channel sounder. (a) TDM-MIMO working principle. (b) Schematic diagram of massive MIMO channel measurement platform.}
	\label{Figure_mimosounding}
\end{figure}

 As illustrated in Fig. \ref{Figure_mimosounding} and Table \ref{mimosounding}, the mid-band massive MIMO platform employs high-speed electronic switches to complete channel measurements within 100 milliseconds, thus facilitating comprehensive TDM-MIMO channel characterization at medium and low frequencies \cite{Miao2024Measurement}. By integrating pseudo-random (PN) sequence signals with time-division multiplexing antenna switching technology, the system effectively decouples the response of the antenna measurement system itself.  $T_t$: transmitting switch duration, the duration of the signal on an array at the transmitting end. $T_r$: receiving switch duration. $T_{cy}$: the time of a probe cycle, which must meet $T_{cy}\geq MT_t$. $T_g$: protection interval, which will be reset to zero in the next period. $T_{sc}$: continuous detection duration of each element at the receiving end. $T_s$: pulse period. In one cycle, each receiving antenna array is detected once. A specified probe duration, $T_t$ contains exactly one detection period, that is, $T_t=NT_{sc}$. As switching takes time in the actual operation process, the protection interval must be set, so the interval between two consecutive detection signals is specified as $T_r$, where $T_r\geq T_{sc}$. 

\par The transmitter end comprises a high-frequency signal generator, control computer, switch matrix, GPS rubidium clock module, feeder, and antenna array. Meanwhile, the receiver end is composed of a MIMO antenna array, feeder, spectrum analyzer, control computer, switch matrix, and GPS rubidium clock module. Figure 1 presents a simplified schematic of the measurement system, including the signal transceiver and the transceiver antenna matrix module.

The transceiver antenna matrix module leverages the GPS reference clock to achieve high-precision synchronization initialization and synchronization unit switching. In addition, the signal transceiver relies on the 10 MHz GPS reference clock to maintain consistency of the transmitter’s base-band signal, the carrier and I/Q demodulation frequencies at the receiver end, and the timing sequence for time-domain sampling. As a result, signal transmission, antenna switching, and signal acquisition all benefit from precise synchronization. The array antenna’s response is measured, demonstrating an angular resolution less than 2°.

\begin{table}[htbp]
	\centering
	\caption{Performance of massive MIMO channel sounding platform covering new mid-band}
	\setlength{\tabcolsep}{0.1 mm}
	\label{mimosounding}
	\renewcommand{\arraystretch}{1.5}
	 \setlength{\tabcolsep}{1 mm}
	\begin{tabular}{c|c}
	 \hline \hline
	    Parameter    & Performance  \\ \hline
		Work frequency [GHz]    & 3-16  \\ \hline
		Max bandwidth [GHz]     & 2  \\  \hline
		Number of Tx switch matrix channel &   128  \\   \hline
		Tx switch gain [dB] &  $\geq$ 27    \\   \hline
		Number of Rx switch matrix channel   & 64  \\   \hline
		Rx switch gain [dB] &  $\geq$ 33      \\   \hline
		Antenna angular resolution [$^\circ$]  &    $\leq$ 2  \\   \hline
		Synchronous mode    & GPS rubidium clock\\   \hline \hline
	\end{tabular}
\end{table}

\section{GROUND COMMUNICATION CHANNEL}
\label{sec:III}

\subsection{Far-Field Channel Characteristics}

\subsubsection{Path Loss Model}
\par Path loss is a fundamental channel model for evaluating the link budget and coverage in cellular network, which is extracted from a set of small areas called local area (LA), where only small-scale fading takes place and multiple measurement positions are included \cite{song2021modeling}. 

\par Due to the large number and diversity of buildings, path loss demonstrates distinct characteristics in both LoS and NLoS scenarios \cite{miao2023Sub-6}.

\begin{figure}[htbp]
	\xdef\xfigwd{\columnwidth}
	\setlength{\abovecaptionskip}{0.1 cm}
	\centering
	\begin{tabular}{cc}
	\includegraphics[width=4.5cm,height=3.5cm]{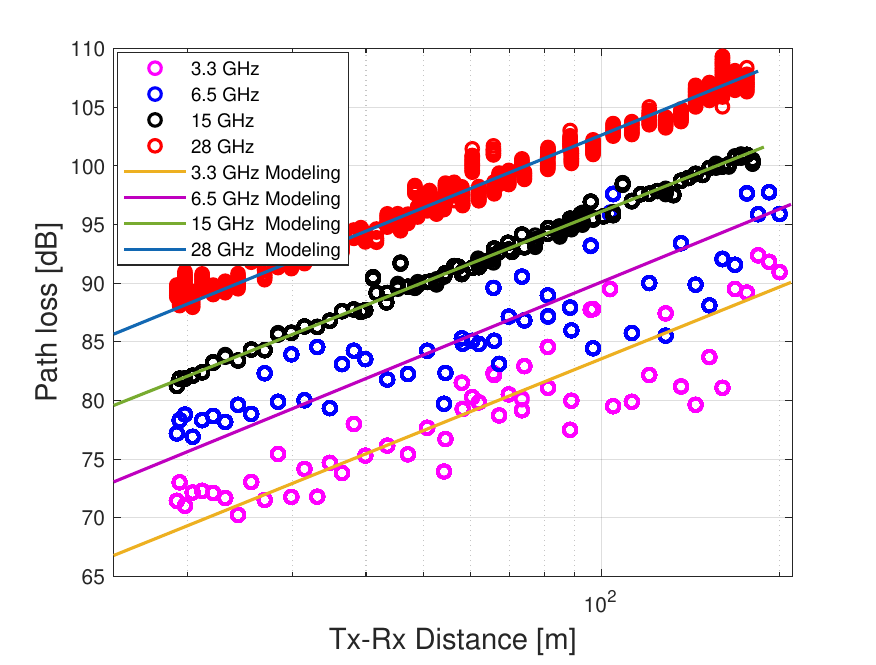} \hspace{-8mm} &\includegraphics[width=4.5cm,height=3.5cm]{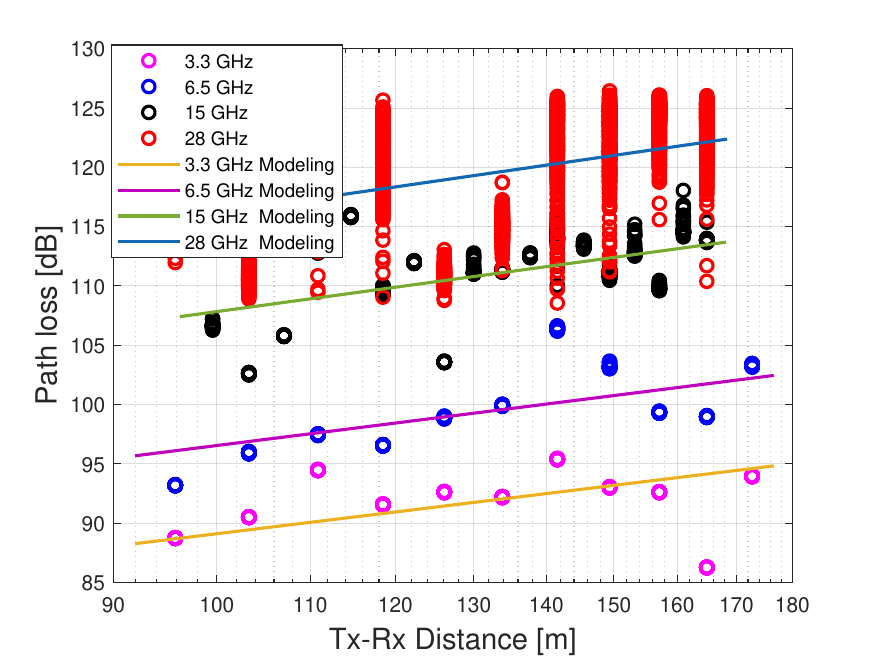}\\
		{\footnotesize\sf (a)} &
		{\footnotesize\sf (b)} \\
	\end{tabular}
	\caption{Path loss model in UMi scenario. (a) LoS. (b) NLoS.}
	\label{path_loss_fig}
\end{figure}

\par  As demonstrated in Fig. \ref{path_loss_fig}, the higher the frequency, the greater the path loss. Moreover, at higher frequencies, path loss becomes more sensitive to environmental changes (e.g., vehicular movement) due to the heightened sparsity of multi-path components (MPCs) . Consequently, snapshots taken at the same measurement location tend to exhibit significant fluctuations. Under NLoS conditions, where a direct LoS path is absent, this effect is even more pronounced. These results also demonstrate that multi-path propagation can help resist deep fading, and abundant multipaths make the fluctuation of path loss smaller.

\par The close-in (CI) free-space reference distance model is employed to integrate path loss and shadow fading into a unified framework

\begin{equation}
\begin{split}
PL(d)& = PL(d_0) + 10n{\rm log}_{10} (\dfrac{d}{d_0}) +X(0,\sigma), \quad d\geq d_0
\end{split} 
\end{equation}
\\ where
\emph{$PL(d_0)$} denotes the closest in free-space path loss (FSPL) in dB, and is a function of wavelength expressed as  $10{\rm log}_{10}(4\pi d_0/\lambda)^2$, with $d_0=1$ m. $n$ is the best fit minimum mean-square error path-loss exponent (PLE) from the simulation results in each case. $X$ is a zero mean Gaussian random variable with a standard deviation $\sigma$ in dB, also known as the shadow factor, representing large-scale signal fluctuations caused by  shadowing by large obstructions in the wireless channel. 

\par In Fig. \ref{path_loss_fig}, the presence of some obstacles (e.g., walls, trees, street light poles, and vehicles) in the environment causes the PLEs in the LoS condition to exceed the FSPL model ($n$ = 2). Nevertheless, these values still align well with the PLE ($n$ = 2.1) of UMi-LoS in 3GPP \cite{3gpp38.901}. A comparison of the PLEs across various LOS conditions and frequency bands reveals that the PLEs are higher in the NLOS condition than in the LOS condition, and that they further increase with rising frequency. This observation indicates that path loss  grows at a faster rate with increasing frequency in NLoS condition.

\par In the LoS condition, although the differences of PLE  across the four frequency bands are minimal, the 15 GHz band exhibits relatively lower shadow fading. In the NLoS condition, the 15 GHz band exhibits relatively lower shadow fading, indicating that the increasing rate of path loss is more comparable to that at the 3.3 GHz band , thereby demonstrating good coverage capability. Furthermore, it is evident that while the coverage capability of 6.5 and 15 GHz is weaker than that at 3.3 GHz, these frequency bands are more capable of achieving stronger signal coverage than the 28 GHz band.

\par In the LoS condition, as the frequency increases, except for the theoretical value, there is no obvious additional path loss fluctuation ($\textless$ 1 dB). Moreover, when the frequency doubles, the path loss model rises by about 6 dB, which can be well verified in the LoS condition. This indicates that the propagation characteristic is very close to that of free space in the LoS condition\cite{miao2023Sub-6}.

\par In the NLoS condition, electromagnetic waves of different frequencies exhibit varying reflection/diffraction/scattering behaviors when encountering obstacles. Consequently, compared to 3.3 GHz, the additional path losses introduced by these higher frequencies are 2.59, 5.98, 8.36 dB, respectively\cite{miao2023Sub-6}. 

\par In our study, the fitting value of $C$ is 9.02 in the NLoS condition\cite{miao2023Sub-6}. The frequency-dependent coefficient (29.02) significantly exceeds the value of 23 reported in WINNER  \cite{WINNERII}. This suggests that the higher frequency results in greater signal loss in the NLoS condition in the UMi environment. Consequently, future systems operating at higher carrier frequencies may need to transmit at higher power levels to maintain coverage over the same area as systems operating at lower frequencies. These findings  indicate that the additional path loss increases with frequency. It can be assumed that the disadvantage of the high-frequency signals in terms of coverage is really prominent. 

\subsubsection{RMS Delay Spread}
\par
\begin{figure}[htbp]
	\setlength{\abovecaptionskip}{0.3 cm}
	\centering
	\includegraphics[width=0.4\textwidth]{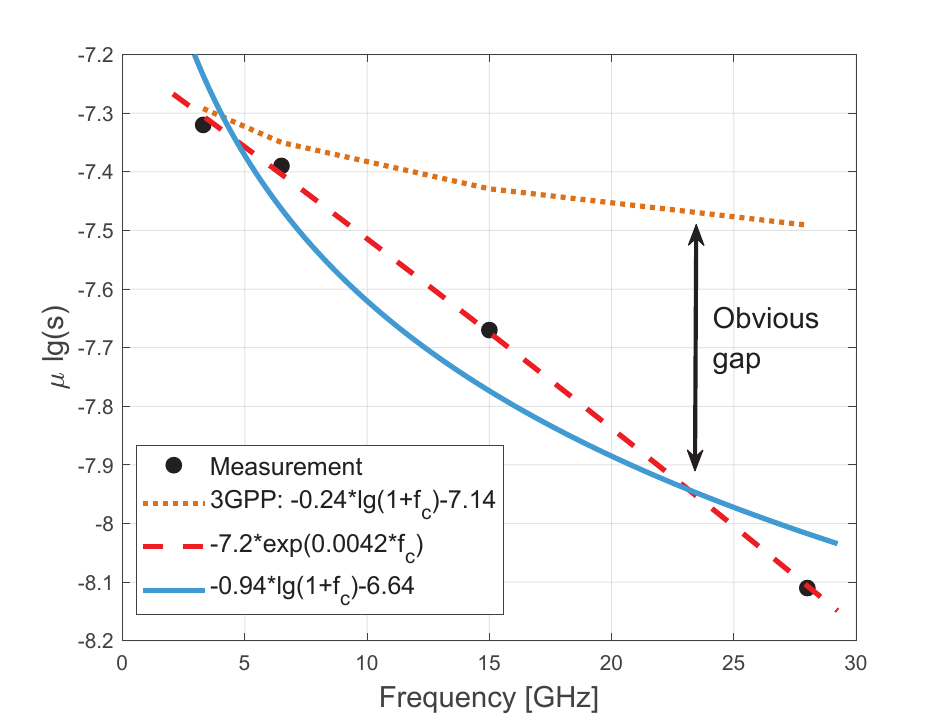}
\caption{Frequency dependence model for delay spread.}
	\label{figure_rms_delay}
\end{figure}

 Fig. \ref{figure_rms_delay} shows the distribution of the RMS delay spread probability density in the 3.3, 6.5, 15 and 28 GHz frequency bands in the UMi scenario.

\begin{table}[htbp]
	\centering
	\caption{Multi-frequency Delay Spread in UMi Scenario}
	\setlength{\tabcolsep}{0.1 mm}
	\label{delay_spread_table}
	\renewcommand{\arraystretch}{1.5}
	\setlength{\tabcolsep}{7mm}
	\begin{tabular}{c|c|c}
		\hline \hline
		Frequency [GHz]    & $\mu$ [ns]   &   $\sigma$ [ns] \\ \hline
		3.3     & 42.77  &  2.03 \\  \hline
		6.5     & 37.93  &  2.12 \\  \hline
		15     & 17.56  &   2.97\\    \hline
		28      &  5.64   & 2.62 \\ \hline
		\hline
	\end{tabular}
\end{table}

\par In Table \ref{delay_spread_table}, it can be observed that higher frequencies correspond to smaller  delay spreads. This occurs because as the frequency increases,  the path loss also increases, leading to significant attenuation of multi-path power. Furthermore, with an increase in carrier frequency, the multi-path becomes less dense and more dispersed. Consequently, the channel is dominated by a small number of effective paths, with the majority of the energies of MPCs concentrated within relatively small regions of the delay domain. The RMS delay spread (about 6 ns) at 28 GHz in the UMi scenario is on the order of a few nanoseconds, which aligns with the measurement results presented in \cite{ying2020analysis}. The average RMS delay spread obtained at low-frequency bands (3.3 GHz and 6.5 GHz) is larger (\textgreater 20 ns). With the relatively more obstacles, it shows that the effective long multi-path is more abundant in the UMi scenario. This result is consistent with the propagation characteristics of low-frequency signals, which exhibit stronger  diffraction capabilities. The UMi scenario contains numerous obstacles, such as trees and buildings. Moreover, the difference of delay spread between the two frequency bands is minimal (\textless 5 ns), suggesting similar multi-path delay characteristics. Consequently, the 5G system frame structure, designed based on the channel characteristics at the 3.5 GHz frequency band, is applicable to the 6 GHz frequency band in the 5G-Advanced system.

\par As illustrated in Fig. \ref{figure_rms_delay}, the frequency-dependent model of delay spread in 3GPP \cite{3gpp38.901} is a statistical model oriented to 0.5-100 GHz. Therefore, the accurate mean value of delay spread can not be obtained in the 6-24 GHz band, which is of particular interest to 5G-Advanced. In our experiments, it is found that modifying the coefficient of the 3GPP model or the exponential model can address this issue.

\par Given that the UMi scenario features more building occlusions, the electromagnetic wave of 15 GHz exhibits both strong diffraction capabilities, similar to those observed in the lower-frequency bands below 6 GHz, and a small delay spread characteristic of higher-frequency bands, such as the mmWave band, which is more advantageous for the design of wireless communication systems. In terms of system design, this implies that higher frequencies allow for a reduction in the required orthogonal frequency division multiplexing (OFDM) guard interval, thereby increasing the transmission rate of useful data.

\subsubsection{RMS Angular Spread}
\par The RMS angular spreads represent the power of MPCs spreading over the angle, serving as a crucial  second-order statistic for characterizing the dispersion of the power angular profile and can be calculated as
 
\begin{equation}
\begin{split}
\psi_{rms}& = \sqrt{\dfrac{\sum_{l=1}^{L}(\psi_l-\psi_{mean})^2P(\psi_{l})}{\sum_{l=1}^{L}P(\psi_{l})}} ,
\end{split} 
\end{equation}
\\ where
$\psi_l$ denotes the azimuth angle $\phi_l$ or elevation angle $\theta_l$, and $\psi_{rms}$ denotes the RMS azimuth angular spread $\phi_{rms}$ or RMS elevation angular spread $\theta_{rms}$. In Table \ref{angular_spread}, ASA, ESA, ASD, and ESD are angular spreads of the azimuth angle of arrival, the elevation angle of arrival, the azimuth angle of departure, and the elevation angle of departure, respectively \cite{Miao2024Measurement}.

\begin{table}[htbp]
	\centering
	\caption{The RMS angular spread in 6 GHz band vs. 3GPP}
	\label{angular_spread}
	\renewcommand{\arraystretch}{1.5}
	\setlength{\tabcolsep}{3mm}
	\begin{tabular}{cc|c|c|c|c}
		\hline \hline
	 \multicolumn{2}{c|}{\multirow{2}{*}{\makecell{RMS \\Angular Spread}}}  &  \multicolumn{2}{c|}{Measurement}  &  \multicolumn{2}{c}{3GPP\cite{3gpp38.901}}\\
		\cline{3-6}
   	&	& \makecell{LoS } &\makecell{NLoS }
		&  \makecell{LoS } &\makecell{NLoS }  \\ \hline
		\multirow{2}{*}{\makecell{ASA}} &$\mu$&1.49&1.65 &1.81&1.87\\ 
		\cline{2-6}
		&$\sigma$ &0.18&0.16&0.20&0.11\\ 
		\hline
		\multirow{2}{*}{\makecell{ASD}} &$\mu$&0.85&1.01 &1.15&1.41\\ 
	\cline{2-6}
	&$\sigma$ &0.26&0.28&0.28&0.28\\ 
	\hline
		\multirow{2}{*}{\makecell{ESA}} &$\mu$&1.26&1.36 &0.95&0.26\\ 
	\cline{2-6}
	&$\sigma$ &0.19&0.13&0.16&0.16\\ 
	\hline
		\multirow{2}{*}{\makecell{ESD}} &$\mu$&1.06&0.96 &-&-\\ 
	\cline{2-6}
	&$\sigma$ &0.30&0.30&-&-\\ 
		\hline	\hline
	\end{tabular}
\end{table}

\par In Table \ref{angular_spread}, it is found that in the 6 GHz band, the mean value of ESD is lower than that of ESA, which aligns with the azimuth angular spread. This can be attributed to the positioning of the Tx antenna on the roof of the building,  where its height exceeds that of the surrounding scatterers, leading to fewer reflection and scattering paths. In contrast, the measurement route contains many scatterers with a height higher than that of the antenna at the Rx end, resulting in more reflection and scattering paths being received.

\subsubsection{Channel Sparsity}

As the frequency increases and wavelength becomes shorter, the loss is generally significant, leading to the sparsity of channels \cite{10452868,9964460}. The channel sparsity is defined as a few dominant MPCs containing most of the power \cite{5454399}. The research of channel sparsity is crucial for minimizing MIMO system overhead and optimizing beamforming algorithms\cite{9679704,9173768}.

Therefore, MIMO channel measurements are conducted at 13 GHz in UMa scenarios to study the channel sparsity. The transmitter (TX) is located on the roof with a height of 27.8 m. The receiver (RX) moves along four lines on the ground, with the lengths of 120, 150, 142, and 150 m, of which three lines are in the LOS condition and one line is in the NLOS condition. The detailed measurement setup parameters are illustrated in Table \ref{parametersparse}. This paper uses SAGE algorithm to estimate the channel parameters to eliminate the influence of antenna response.


\begin{table}[!tbp]
\renewcommand\arraystretch{1.5}
\caption{Measurement Parameters}
\begin{center}
\begin{tabular}{p{2.2cm}<{\centering}|p{4cm}<{\centering}}
\hline
\hline
\textbf{Parameters}&\textbf{Value/Type} \\
\hline
Scenario&UMa \\
\hline
Carrier Frequency& 13 GHz \\
\hline
RF Bandwidth&400 MHz \\
\hline
TX Antenna Type & UPA with 128 elements \\
\hline
RX Antenna Type & ODA with 64 elements \\
\hline
Polarization &$\pm 45^{\circ} $\\
\hline
TX Signal &PN9 \\
\hline
\hline
\end{tabular}
\label{parametersparse}
\vspace{-1.2em}
\end{center}
\end{table}

The Gini index is widely used to measure the level of sparsity because the definition of the Gini index is similar to the channel sparsity, expressed as \cite{Gini}

\begin{equation}
G=1-2\sum_{i=1}^{R}\frac{p_i}{\left \| \mathbf{p}   \right \|_1 }  \left ( \frac{R-i+\frac{1}{2} }{R}  \right ), 
\label{Gini}
\end{equation}

\noindent where $p_i$ represents the power of the $i$-th ray. $\mathbf{p}$ represents a power vector composed of the power of $R$ rays. The elements in $\mathbf{p}$ are arranged in an ascending order. The subscripts are after sorting. The range of Gini index is $[0,1]$. The closer the Gini index is to 1, the sparser the channel is.

\begin{figure}[!tbp]
\centering
\includegraphics[scale=0.5]{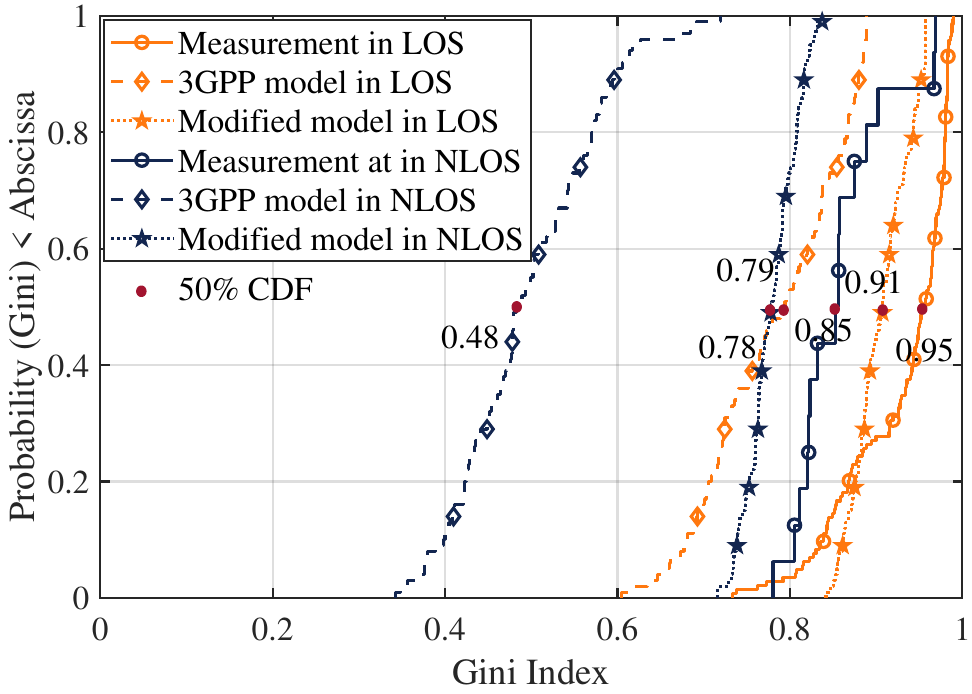}
\caption{The CDF of Gini index generated by measured channels, by the 3GPP channel model, and by the 3GPP model with proposed intra-cluster power allocation model (modified model). (a) LoS scenario. (b) NLoS scenario.}
\label{GiniCDF} 
\end{figure}

The channel sparsity in the UMa scenario is illustrated by the cumulative density function (CDF) of the Gini index in Fig. \ref{GiniCDF}. The results of MIMO channel measurements in both LOS and NLOS conditions in the UMa scenario show that the channels at 13 GHz is sparse. Specifically speaking, the Gini index in LOS condition is 0.95 at 13 GHz. In NLOS condition, the Gini index is 0.86 at 13 GHz. It is found that the Gini index in NLOS condition is smaller than that in LOS condition because the power of the LOS components contains much power. The dominant rays in NLOS condition are specular reflection rays and first-order diffraction rays.

According to the 3GPP channel coefficient equation, the power of each ray within a cluster is generated based on the principle of evenly dividing the power of the cluster, as expressed as

\begin{equation}
H_{n} \left ( t \right ) =\sqrt{\frac{P_n}{M} } \sum_{m=1}^{M} c_{m}\exp \left ( j2\pi v_{m}t  \right ),
\label{h}
\end{equation}

\noindent where $P_n$ represents the power of the $n$-th cluster, $M$ represents the number of rays per cluster. $c_{m}$ is the complex channel coefficient. The coefficient includes the radiation pattern of the antennas, random phases and path loss. $v_{m}$ is the Doppler frequency.

The dashed lines with diamond in Fig. \ref{GiniCDF} show that the Gini index obtained from the 3GPP channel model is generally smaller than the measurement results. The Gini index of 3GPP model in NLOS condition is 0.49 for the $50\%$ cases of the CDF, which is 0.37 smaller than the measurement in the same situation. It is worth noting that we used the measured K-factor in 3GPP model. Therefore, it is indicated that the 3GPP channel model cannot accurately characterize channel sparsity in both LOS and NLOS conditions of the UMa scenario. This conclusion is consistent with the indoor scenario, thus this paper considers modifying the 3GPP channel model using the intra-cluster power distribution method mentioned in \cite{10452868} to address this deficiency. 

By introducing a parameter, Intra-cluster Power factor (ICP), the distribution of power within a cluster is altered such that there exists a dominant ray within a cluster containing most of the cluster power. The ICP is obtained by analyzing the distribution of the measured power distribution. The modified 3GPP channel coefficient is expresses as

\vspace{-5pt}
\begin{align}
\label{hk}
H_{n} \left ( t \right )& =\sqrt{\frac{ICP\cdot P_n}{ICP+1}} c_{1}\exp \left ( j2\pi v_{1}t  \right )\\
&+\sqrt{\frac{P_n}{\left ( ICP+1 \right) \left ( M-1 \right )}}\sum_{m=2}^{M} c_{m}\exp \left ( j2\pi v_{m}t  \right ).
\notag
\end{align}

\noindent with

\begin{equation}
\vspace{-0.5em}
ICP=\frac{\max\left ( \mathbf{p}_n  \right ) }{\left \| \mathbf{p}_n \right \|_1 -\max\left ( \mathbf{p}_n  \right )}, 
\label{Kc}
\end{equation}

\noindent where $\mathbf{p}_n$ represents the vector composed of the power of all rays within the $n$-th cluster.

The dotted lines with pentagram in Fig. \ref{GiniCDF} show that the Gini index curves obtained by the modified 3GPP channel model with ICP are closer to the measurement results. In the NLOS condition, the $50\%$ cases of the CDF show that the Gini index obtained from the modified 3GPP channel model is 0.79 at 13 GHz, which is differ from the measurement results by 0.07. Compared with the Gini index obtain from 3GPP channel model in the same situation, the accuracy improved by about 5 times. In the LOS condition, the Gini index of the modified 3GPP channel model is 0.89 for the $50\%$ cases of CDF at 13 GHz, which differ from the measurement results by 0.06. But the Gini index of the 3GPP channel model is 0.79 for the $50\%$ cases of CDF at 13 GHz, that is, the modified model reduced the gap between the model and the measurements to one-third in the same situation. These results indicate that the modified 3GPP channel model \cite{3gpp38.901} characterizes the channel sparsity more accurately.

\subsubsection{Channel Capacity}

\par
\begin{figure}[htbp]
	\xdef\xfigwd{\columnwidth}
	\setlength{\abovecaptionskip}{0.1 cm}
	\centering
	\begin{tabular}{cc}
		\includegraphics[width=4.5cm,height=3.5cm]{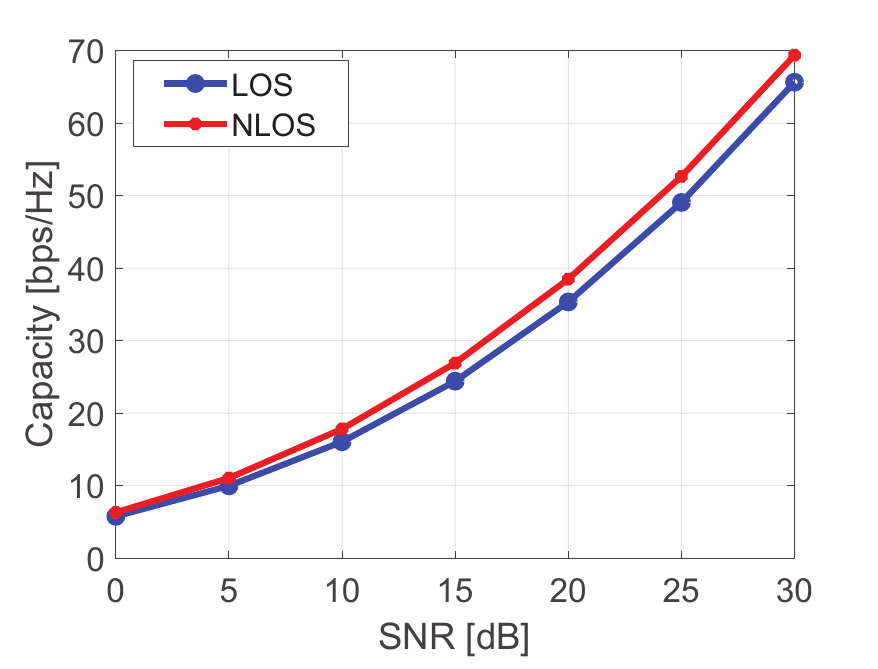} \hspace{-8mm}
		& \includegraphics[width=4.5cm,height=3.5cm]{figure6-a-eps-converted-to.pdf}\\
		{\footnotesize\sf (a)} &	{\footnotesize\sf (b)} \\	
	\end{tabular}
	\caption{The channel capacity in 6 GHz band with 200 MHz bandwidth. (a) 16 Rx antenna elements. (b) 56 Rx antenna elements.}
	\label{fig:Figure6_capacity}
\end{figure}

 Fig. \ref{fig:Figure6_capacity} gives the influence of the environments and number of elements on channel capacity at 6 GHz band in UMa scenario\cite{Miao2024Measurement}.

\par In a NLOS environment, the channel capacity is generally higher compared to a Line-of-Sight (LOS) environment. This is primarily due to the increased complexity of the NLOS environment, where the multipath components exhibit greater degrees of freedom in the angle domain. Additionally, as the number of antenna array elements increases, the channel capacity shows a significant improvement, as illustrated in the Fig. \ref{fig:Figure6_capacity}. Notably, with a higher number of antenna elements, the channel capacity grows more rapidly with increasing signal-to-noise ratio (SNR).

\par These findings highlight that increasing the number of antenna elements is one of the most effective methods to enhance channel capacity. This insight is a key reason why XL-MIMO is emerging as a potential cornerstone technology for 6G communication systems. XL-MIMO leverages much larger-scale antenna arrays, and with further expansion of the antenna scale, it promises to deliver even higher throughput and spectral efficiency, making it a critical enabler for future wireless communication systems.

\begin{figure}[htbp]
\centering
\includegraphics[width=1.0\linewidth]{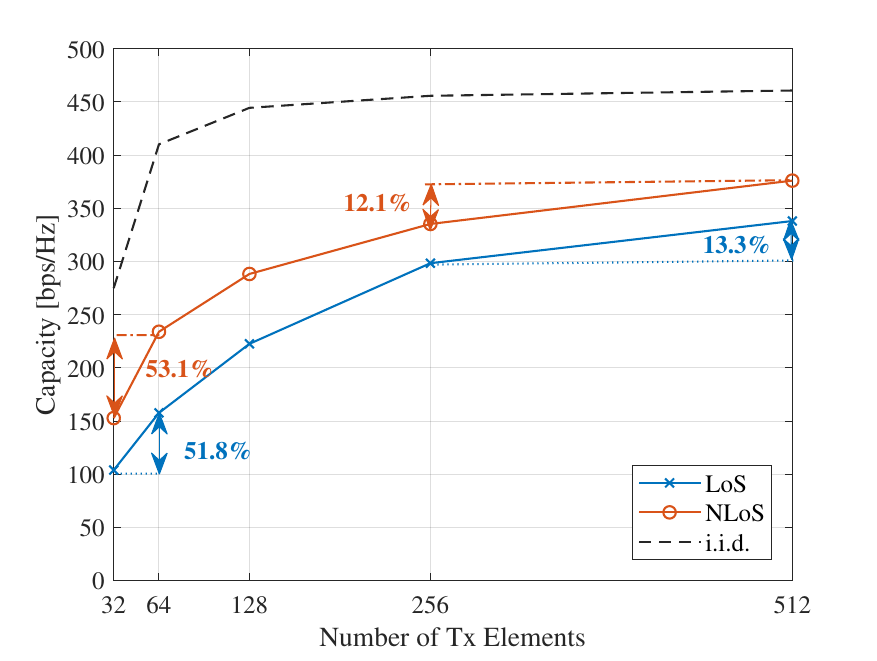}
\caption{Channel capacity for various number of elements, SNR=25dB.}
\label{Capcompare}
\end{figure}

In order to compare the indoor environment, we also observe the capacity characteristics of the corridor scenario\cite{Wei2024Corridor}. Fig. \ref{Capcompare} illustrates how the channel capacity varies with the number of transmitting antenna elements increasing from 32 to 512. The results indicate that the growth rate of channel capacity decreases as the number of transmitting antenna elements is sequentially doubled, from 51.8$\%$ to 13.3$\%$ for LoS and from 53.1$\%$ to 12.1$\%$ for NLoS. The channel capacity under i.i.d. channel does not show a substantial increase  when the number of transmitting antenna elements exceeds 64. This is due to the limited number of 56 antenna elements at the receiver end, causing the capacity to be saturated when the number of transmitting antenna elements surpasses this threshold.

Furthermore, there exists a relatively significant disparity between the channel capacity under the corridor scenario and that under the i.i.d. channel. Specifically, the channel capacity in the LoS and NLoS environments achieves a maximum of 73.4$\%$ and 81.6$\%$ of the i.i.d. channel capacity, respectively. This difference arises due to the relatively closed space of the corridor scenario, where the spatial freedom of the MPCs is limited. The research presented in Fig. \ref{Capcompare} indicates that XL-MIMO performs comparably  to the i.i.d. channel in open environments with relatively abundant scattering.

\subsection{Near-Field Effects and Spatial Non-Stationary}

\subsubsection{Spherical-Wave Property}
\par
\begin{figure}[htbp]
	\xdef\xfigwd{\columnwidth}
	\setlength{\abovecaptionskip}{0.1 cm}
	\centering
	\begin{tabular}{cc}
		\includegraphics[width=4.2cm,height=3cm]{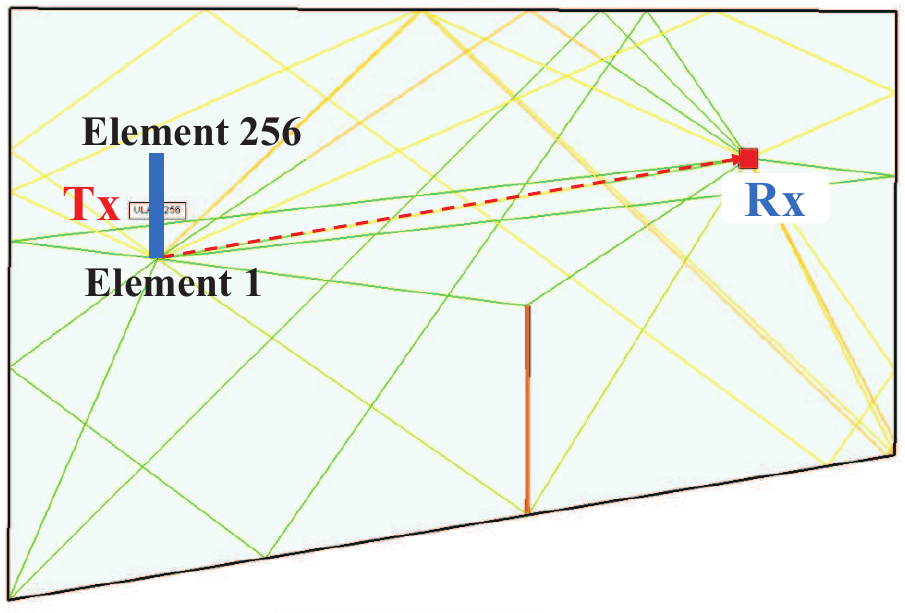} \hspace{-5mm} &\includegraphics[width=4.5cm,height=3.2cm]{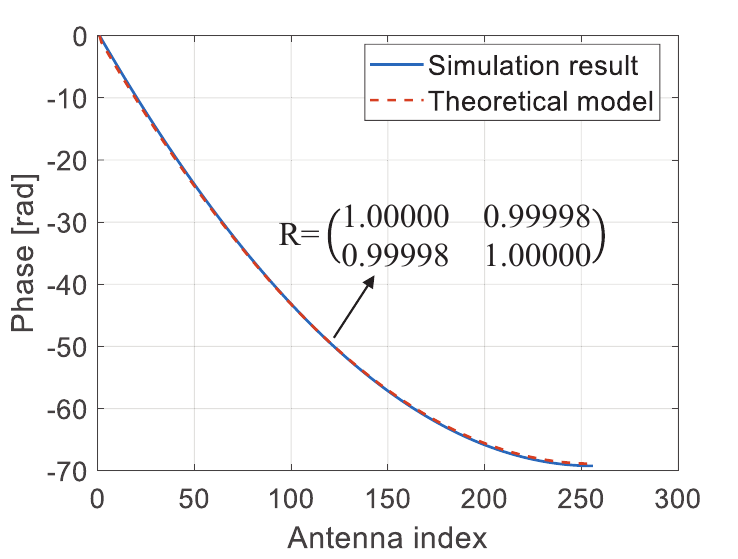}\\
		{\footnotesize\sf (a)} &
		{\footnotesize\sf (b)} \\
	\end{tabular}
	\caption{The phase verification. (a) Simulation scenario. (b) Simulation results and theoretical model.}
	\label{fig:Nearfield}
\end{figure}

 The spherical-wave signal model's accuracy is demonstrated through a Wireless Insite simulation scenario, where the initial 256-element array antenna is depicted in Fig. \ref{fig:Nearfield}(a). The phase simulation results and the theoretical results, derived using the spherical-wave signal model, are presented in Fig. \ref{fig:Nearfield}(b).

\par In Fig. \ref{fig:Nearfield}(b), the phase derived from the simulation closely aligns with the established signal model\cite{miao_near}. By applying correlation to compare the two sets of values, it is evident that the correlation between them is remarkably strong, suggesting that the theoretical model exhibits excellent accuracy.

\subsubsection{Power Attenuation Factor}
The author in\cite{gao20233gpp} presents an spatial non-stationary (SnS) channel model framework that builds upon the 3GPP model outlined in \cite{3gpp38.901}. This framework integrates SnS characteristics, extending the 3GPP 3-dimensional (3D) GBSM method to represent near-field propagation and SnS channels. The channel coefficient for the $n$-th cluster between the $s$-th BS antenna element and the UT is given by:

\begin{equation}\label{equ1}
\mathbf{H}_{n,s}^{SnS}(t)=\mathbf{S}_{n,s}\cdot\mathbf{H}_{n,s}^{3GPP}(t)\cdot\mathbf{A}_{n,s},
\end{equation}
where the UT is assumed to have a single antenna. The channel coefficient $\mathbf{H}_{n,s}^{3GPP}(t)$ includes the antenna radiation pattern, random initial phases, and Doppler phase, with calculations based on the 3GPP model as outlined in \cite{3gpp38.901}. The entry $\mathbf{A}_{n,s}$ of the manifold matrix $\mathbf{A}$ simulates phase and power variations resulting from spherical propagation, assuming near-field conditions. $\mathbf{S}_{n,s}$ is the $(n,s)$-th entry of the SnS matrix $\mathbf{S}$, which is utilized to simulate the visibility of the cluster and is defined as:

\begin{equation}\label{equ2}
\mathbf{S}_{n,s}=\begin{cases}1,&\quad observed\\0,&\quad not\ observed\end{cases}
\end{equation}

In this model, cluster visibility is determined by the visible region (VR) of the antenna elements, as defined in \cite{liu2012cost}. However, the model only addresses whether clusters are observed along the antenna elements, influenced by SnS characteristics, without accounting for power variation of clusters within the VR that arises from these characteristics \cite{yuan2022spatial}.

The power attenuation factor $\alpha_{n,s}$ is introduced to model the power variation of clusters within the visibility region of the antenna elements, where $\alpha_{n,s} \in (0,1]$. The revised expression for $\mathbf{S}_{n,s}$ is given by:

\begin{equation}\label{equ3}
\mathbf{S}_{n,s}=\alpha_{n,s}\cdot\begin{cases}1,&\quad observed\\0,&\quad not\ observed\end{cases}
\end{equation}

The power attenuation factor, denoted as $\alpha_{n,s}$, is extracted from the path characterized by significant power variations and is defined as follows:

\begin{equation}\label{equ4}
\alpha_{n,s}=\frac{P_{n,s}}{P_{n}^{\max}},
\end{equation}
where $n$ represents the $n$-th path present in the channel measurement, and $s$ corresponds to the $s$-th MIMO antenna element at the Tx. The variable ${P_{n,s}}$ denotes the power of the $n$-th path from the $s$-th antenna element at the Tx to the Rx, while ${P_{n}^{\max}}$ refers to the maximum power of the $n$-th path from all antenna elements at the Tx to the Rx. The factor $\alpha_{n,s}$ lies within the interval $(0,1]$.

\begin{figure}[h]
\centering
\subfloat[]{\includegraphics[width=0.25\textwidth]{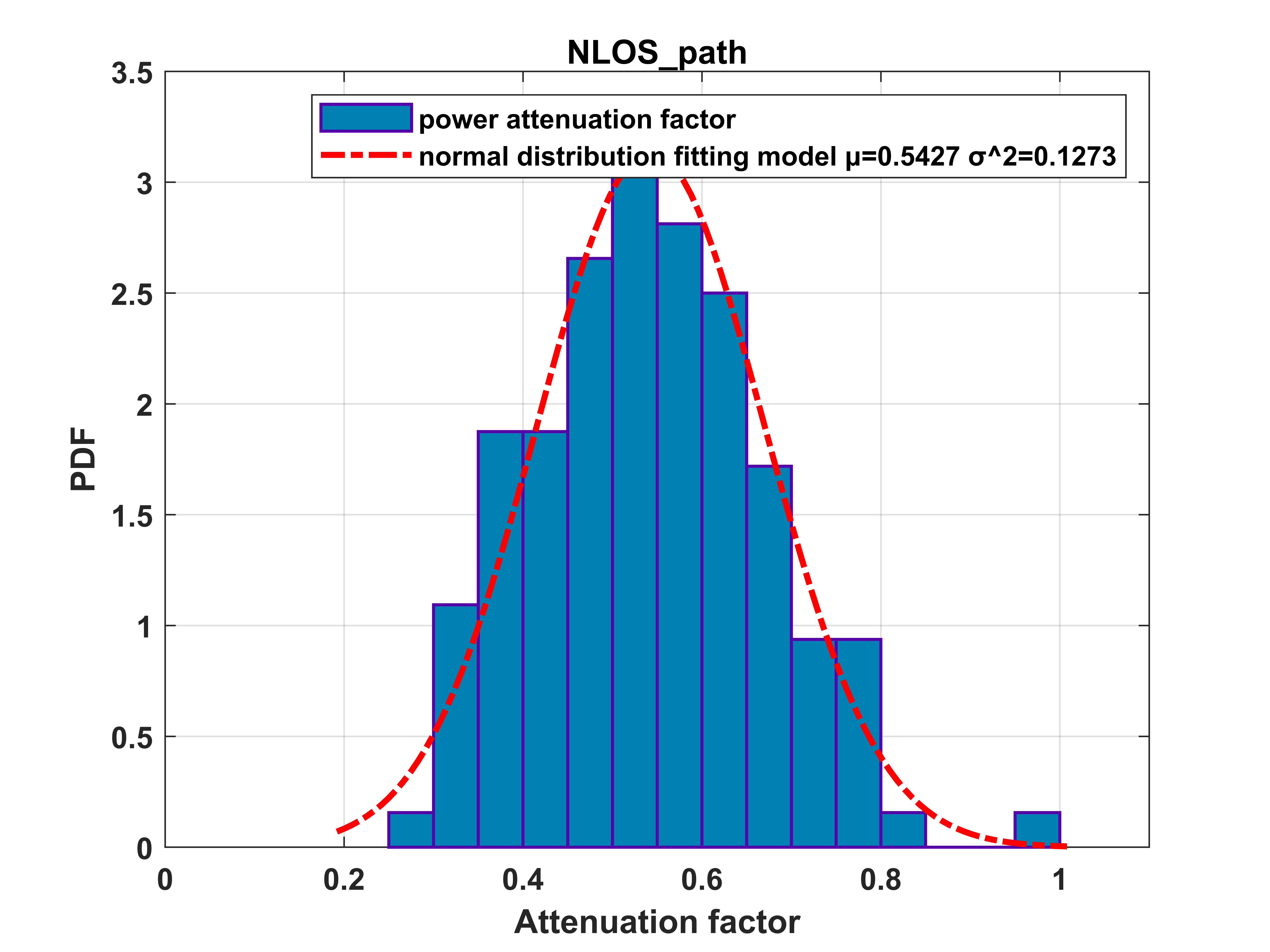}}\label{fig:left figure}%
\subfloat[]{\includegraphics[width=0.25\textwidth]{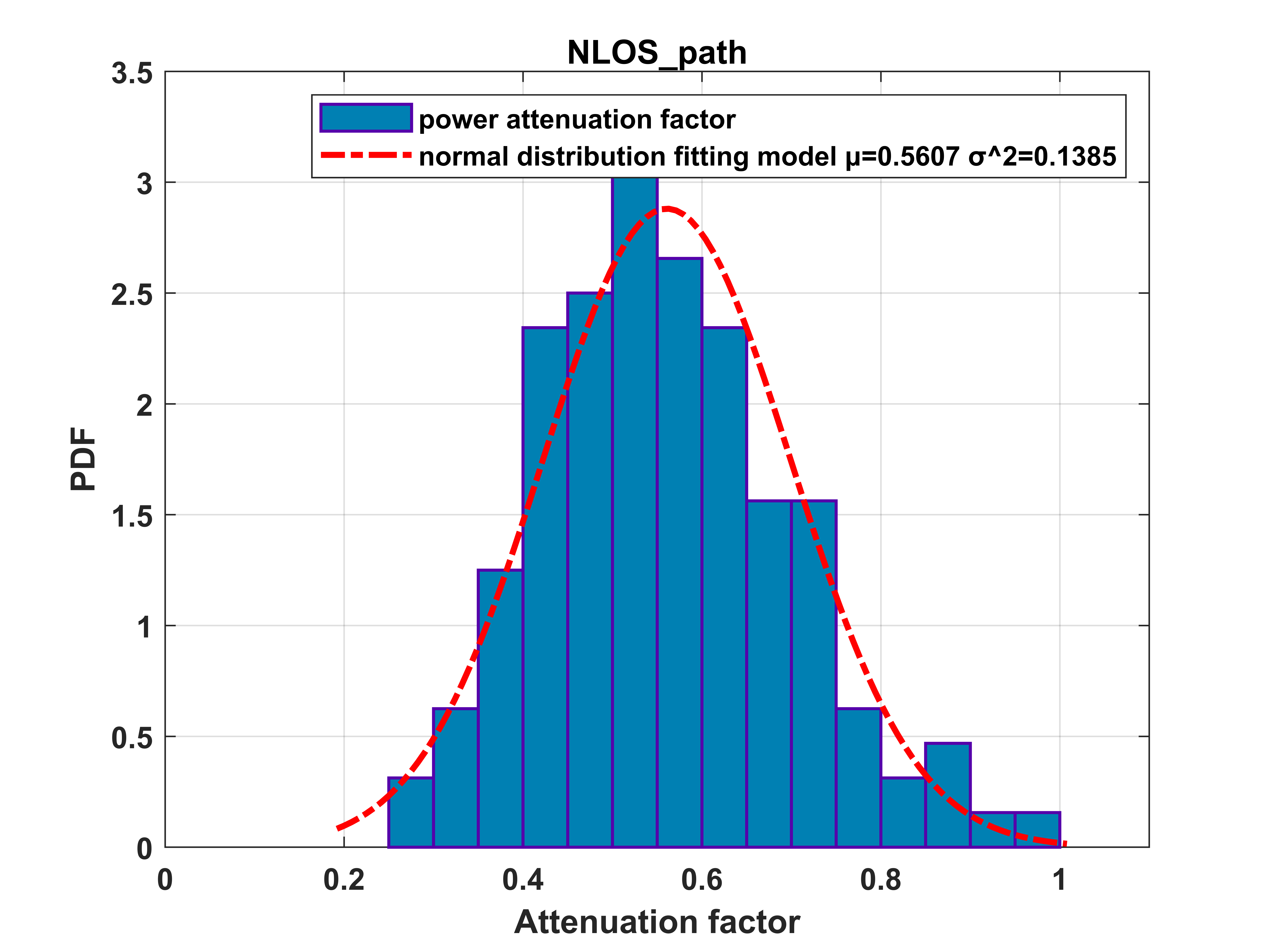}}\label{fig:right figure}%
\caption{Path power attenuation factor $\alpha_{n,s}$ distribution.}
\label{fig:distribution}
\end{figure}

The power attenuation factors of the paths are derived from the measurement data using Equation~\ref{equ4}. As shown in Fig.~\ref{fig:distribution} , the probability density function (PDF) of the power attenuation factors $\alpha_{n,s}$  is presented across different antenna elements at the Tx, accompanied by the corresponding fitting results. The analysis reveals that these power attenuation factors closely follow the normal distribution.

To model the power attenuation factor $\alpha_{n,s}$, the maximum path power ${P_{n}^{\max}}$ (where $n = 1, 2, \ldots, N$) is normalized and the expression is given by:

\begin{equation}\label{equ5}
P_{n}=\frac{P_{n}^{max}}{\Sigma_{n=1}^{N}P_{n}^{max}},
\end{equation}
where we assume there are $N$ paths in the channel measurement corresponding to the $N$ clusters defined in the 3GPP model. $P_{n}$ represents the normalized power of the $n$-th cluster, calculated in step 6 of the 3GPP channel modeling procedure \cite{3gpp38.901}.

\begin{figure}[h]
\centering
\subfloat[]{\includegraphics[width=0.25\textwidth]{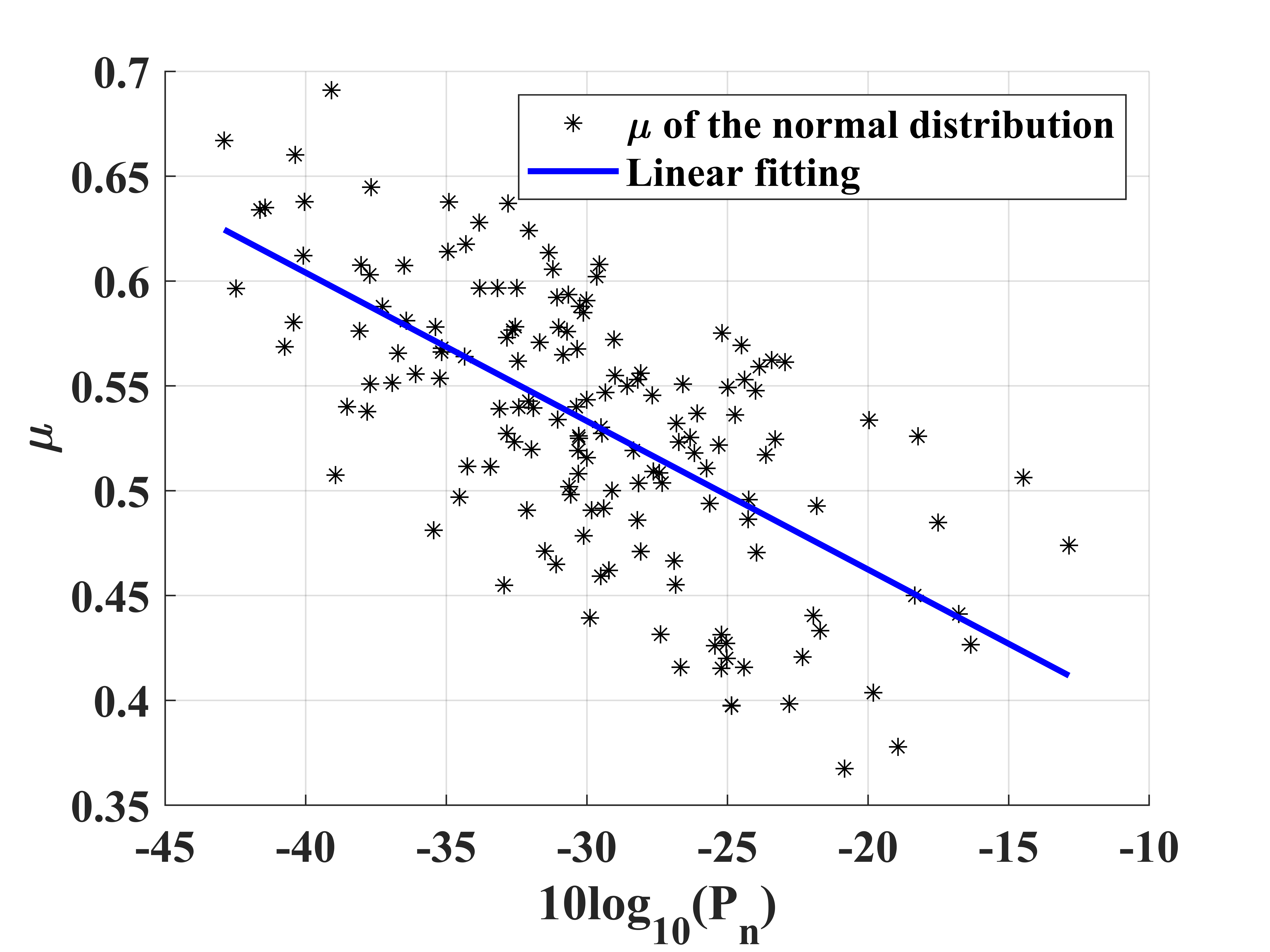}%
\label{fig:left figure}}
\subfloat[]{\includegraphics[width=0.25\textwidth]{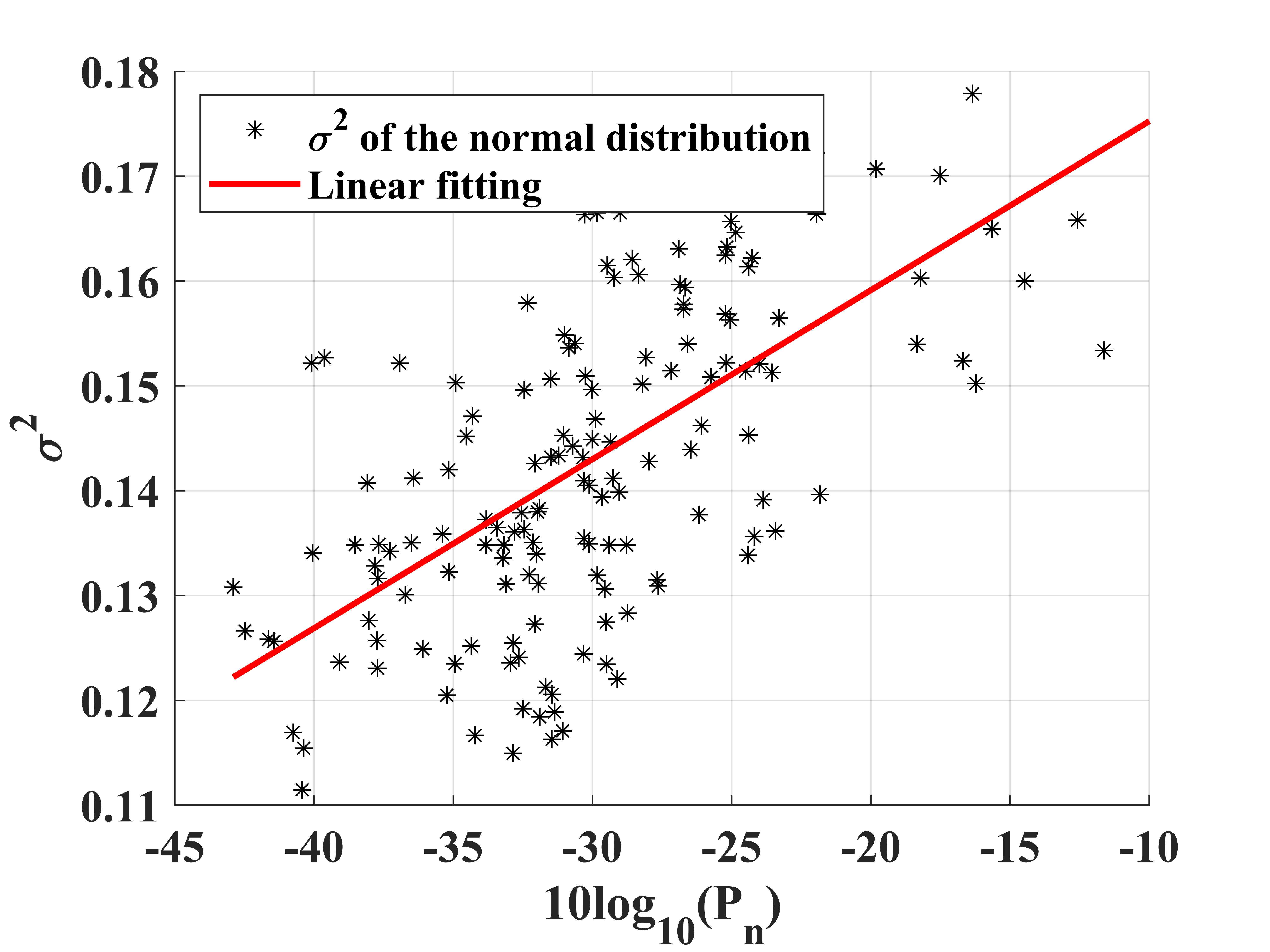}%
\label{fig:right figure}}
\caption{The relationship between the $\mu$, $\sigma^2$ and normalized power $P{_n}$.}
\label{fig:linear}
\end{figure}

The mean ($\mu$) and variance ($\sigma^2$) of the normal distribution that the power attenuation factor $\alpha_{n,s}$ follows are then extracted. The relationship between these parameters and the normalized power $P_{n}$ is depicted in Fig. \ref{fig:linear}.
It can be observed that the mean ($\mu$) and variance ($\sigma^2$) of the power attenuation factor $\alpha_{n,s}$, which is normally distributed, show a linear relationship with the normalized power $P_{n}$. The linear relationship can be described by the following fitting function:
\begin{equation}\label{equ6}
\mu_n(p_n)=10a{_\mu}\text{log}_{10}(p_n)+b_\mu,
\end{equation}
\begin{equation}\label{equ7}
{\sigma^2}{_n}(p_n)=10a_{\sigma^2}\text{log}_{10}(p_n)+b_{\sigma^2}.
\end{equation}
The characteristics of the power attenuation factor $\alpha_{n,s}$  derived from the previous analysis are applied to 3GPP channel modeling. This application yields the power attenuation factor $\alpha_{n,s}$ for the clusters along the BS antenna elements.

\subsubsection{Stationary Interval Division}

To accurately model the SnS phenomenon, \cite{zwr1} proposes a method for stationary sub-interval partitioning based on channel characteristics. This approach assumes that the channel remains stationary within each sub-interval, but exhibits non-stationary across different intervals. The method comprehensively considers factors such as channel correlation, delay spread (DS), azimuth angle spread of departure (ASD), and the birth-death behavior of MPCs for sub-interval partitioning. By analyzing the independence of sub-intervals, this study demonstrates that the proposed method outperforms the conventional averaging technique in terms of sub-interval partitioning.

  


\par For the ULA, we can first obtain the MPCs parameter data $X$, $X = \{x_{1}, x_{2}, \ldots, x_{K}\}$ based on RT simulation, where $x_{1}$ represents the MPCs parameters of the first array element, including power, delay, phase, and angle. Then we can calculate the channel impulse response (CIR) for each transmitting element, which is used for subsequent characteristics analysis.

After processing the parameters\cite{zwr1} , we can proceed with the partitioning of stationary intervals based on the characteristic analysis in Section III. Initially, we preselect interval nodes based on MPCs birth-death, with the selection method outlined by the following expression
\begin{align}
S(k) = \begin{cases}
    1 & \text{if } p_{k}^{n} - p_{k-1}^{n} \ge 3\, \text{dB} \\
   -1 & \text{if } p_{k}^{n} - p_{k-1}^{n} \le -3\, \text{dB} \\
    0 & \text{else}\\
\end{cases}
 ,
\end{align}

\noindent where $k$ belongs to the range 1 to $K$, $K$ is the number of transmitting array elements, $n$ belongs to the range 1 to $N$, $N$ is the total number of MPCs from the $k$-th array element. The value of $S(k)$ represents the birth and death situation at the $k$-th element, $S(k) = 1$ indicates the birth of a new MPC at the $k$-th element, $S(k) = -1$ indicates the death of a MPC at this location, and $S(k) = 0$ indicates no birth or death phenomenon at the $k$-th element. The specific selection criterion is based on the power variation of multipath signals on the array domain.
$p_k^n$ represents the power of the $n$-th path at the $k$-th array element, and $p_{k-1}^n$ represents the power of the $n$ path at the $(k-1)$-th array element. If there exists a path for which $p_k^n-p_{k-1}^n \ge$ 3 dB, then set $S(k)=1$, when $p_{k}^n-p_{k-1}^n \le$ 3 dB, set $S(k)=-1$. By iterating through all elements, we obtain the complete $S$ vector.

After obtaining the $S$ vector, further refinement of the selection of interval nodes is necessary, utilizing channel correlation, DS, and ASD. By analyzing the variations in the parameters at the array elements, elements from $S$ are identified as potential interval boundaries. 
First, by analyzing the changes in correlation, if a certain interval can be considered stationary, the correlation coefficients for sub-channels within that interval will be relatively large and exhibit minimal fluctuations. When a new element is introduced to this interval, a significant increase in fluctuations may indicate that the newly added element serves as a boundary. The fluctuations are calculated using the Mean Absolute Deviation (MAD).

\begin{align}
M A D=\frac{1}{n} \sum_{i=1}^{n}\left|x_{i}-\bar{x}\right|,
\end{align}
where $n$ is the number of the data, $x_{i}$ represents the $i$-th data, and $\bar{x}$ is the mean of all data.
A similar approach can be applied to DS and ASD. If adding an element to an interval leads to a significant  rise in parameter fluctuations, then that element may also be a boundary element. Jointly considering the channel correlation, DS, and ASD, the following equation is utilized for evaluation.

\begin{align}
w_{c} \cdot \frac{M_{k}^{C}}{M_{k-1}^{C}}+w_{a} \cdot \frac{M_{k}^{A}}{M_{k-1}^{A}}+w_{d} \cdot \frac{M_{k}^{D}}{M_{k-1}^{D}}>\rho ,\label{eq:lianhe}
\end{align}

\noindent where $M_{k}^{C}$ represents the fluctuation of channel correlation within the interval when the $k$-th element is chosen as the boundary, and $M_{k-1}^{C}$ represents the fluctuation when the $(k-1)$-th element is taken as the boundary. 
Similarly, $M_{k}^{A`}$ and $M_{k}^{D}$ measure the fluctuation of ASD and DS, respectively. 
$w_{c}$, $w_{a}$ and $w_{d}$ represent the weights assigned to the fluctuation values of correlation coefficient, ASD, and DS, respectively. 
$\rho$ is the threshold for judgment. 
The inequality \eqref{eq:lianhe} is used to evaluate each element in $S$, and if the calculated result at a certain element exceeds the threshold, that element is considered a boundary node. After completing the traversal, the final interval partitioning result $Z$ is obtained.

\begin{figure}[htbp]
\centerline{\includegraphics [width=0.45\textwidth] {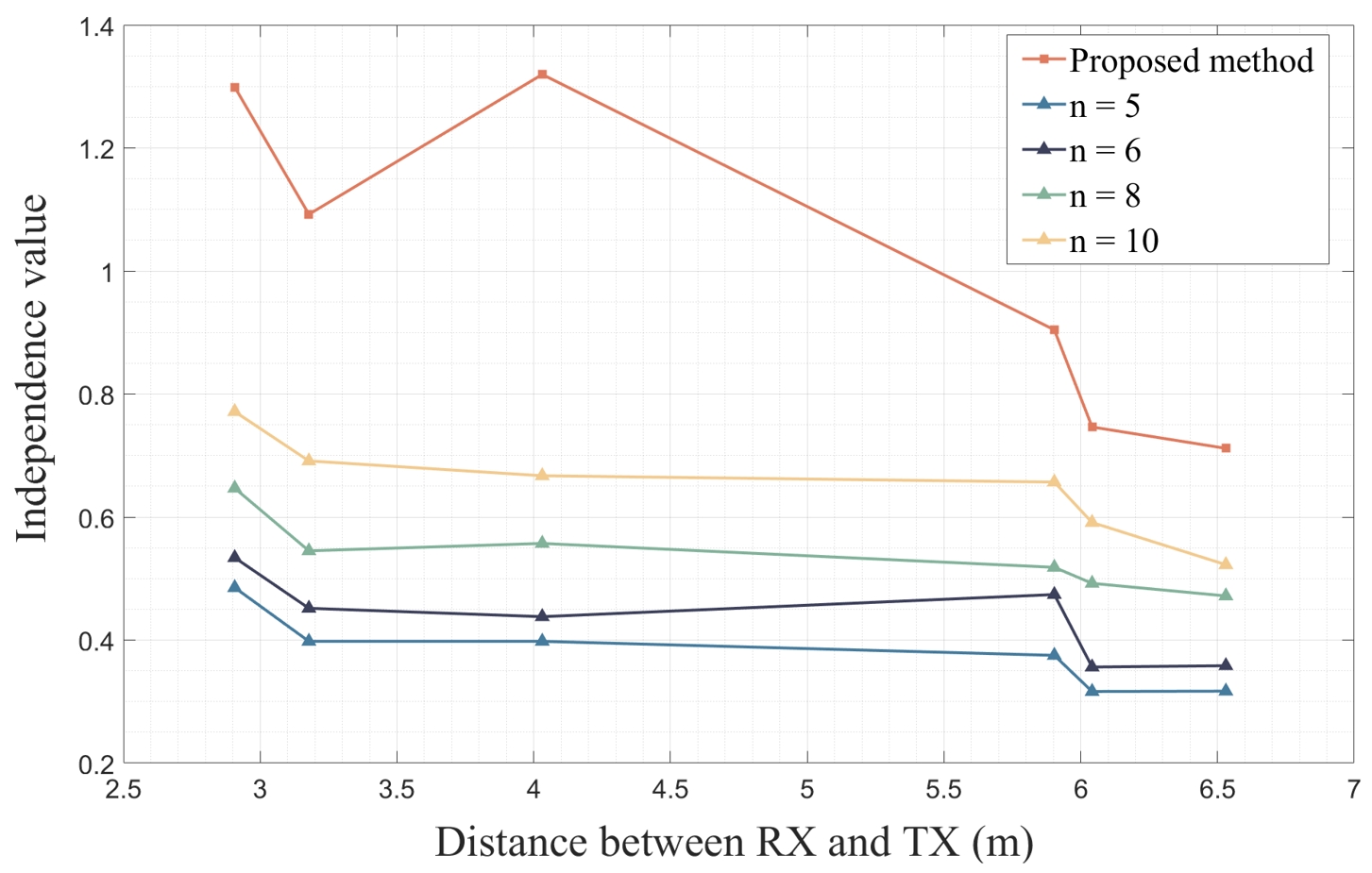}}
\caption{The degree of independence of sub-intervals obtained by different partitioning methods. The $n$ represents the number of array elements in each sub-interval during average partitioning.}
\label{fig:duli}
\end{figure}

After completing the partitioning, the reconstructed CIR for each sub-interval can be revived. Then applying  equation \eqref{eq:mubiao}, we analyzed the independence of the sub-intervals obtained based on channel characteristics and average partitioning. The results are presented in Fig.~\ref{fig:duli}.

\begin{align}
D =\frac{1}{N \cdot (M-1)} \sum_{k=1}^{M-1} \sum_{i=1}^{N}\left|h_{k+1}(i)-h_{k}(i)\right|,\label{eq:mubiao}
\end{align}
where $N$ and $M$ respectively represent the number of delay bins and sub-intervals, $h_{k}$ represents the reconstruction CIR of the $k$-th sub-interval.

In Fig.~\ref{fig:duli}, the horizontal and vertical axes correspond to the Tx-Rx distance and independence level, respectively. 
It is evident that since traditional partitioning techniques neglect environmental factors, the independence of sub-intervals is influenced by the size of the intervals, thereby resulting in poor robustness.
Conversely, when partitioning is based on channel characteristics, the method accurately assigns elements with different scattering environments to distinct intervals. This results in enhanced accuracy and robustness in the obtained outcomes.

Additionally, it is observed that as the distance between the transmitter and receiver increases, the independence of the sub-intervals for different methods decreases. This occurs because with the distance surpassing the Rayleigh distance, the scattering environments in different sub-intervals become identical, the SnS phenomenon becomes less pronounced, and the independence of the sub-intervals becomes zero.

\section{AIR-to-GROUND COMMUNICATION CHANNEL}
\label{sec:IV}

Clutter loss is an important part of the loss in satellite-ground link transmission, which is caused by the occlusion and scattering caused by the buildings near the ground receiving terminal and the natural environment. This section will discuss the relationship between clutter loss and frequency within the FR3 band.

\subsection{Measurement Setting}

The R$\&$S time-domain channel sounder is employed for conducting channel measurements, comprising both  a Tx side and a Rx side. At the Tx side, a vector signal generator (R$\&$S SMW 200A) is utilized to modulate a PN sequence through binary phase shift keying (BPSK). The modulated signal undergoes filtering by a root-raised cosine filter with roll-off factor of 1. At the Rx side, the received signal is demodulated and captured using a frequency spectrum analyzer (R$\&$S FSW 43). To enhance the signal-to-noise ratio (SNR) and dynamic range, a power amplifier (PA) is used for the RF signal during transmission and a low-noise power amplifier (LNA) is employed for signal reception. 

\begin{figure}[htb]
    \centering
    \includegraphics[width=0.45\textwidth]{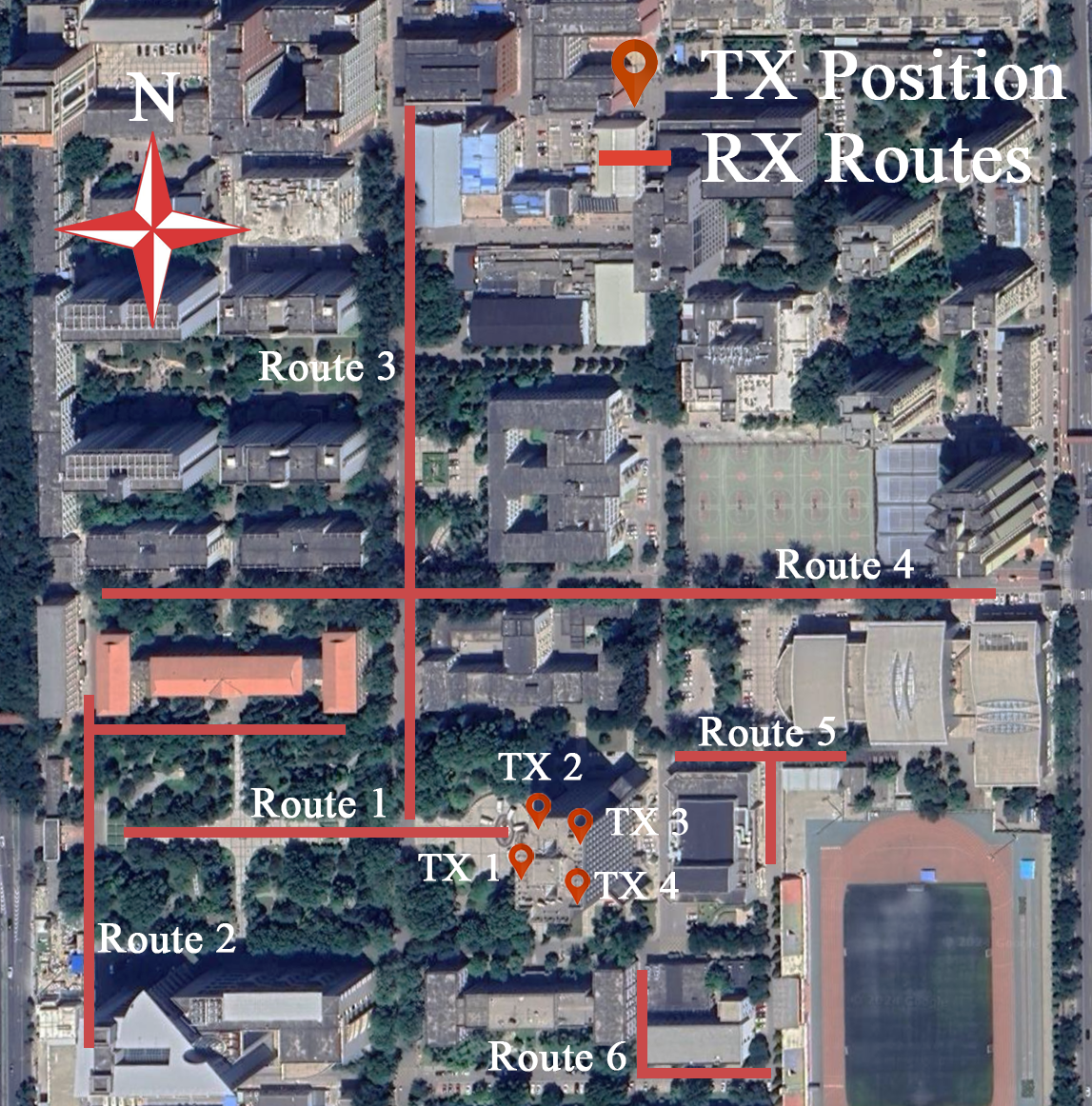}
    \caption{Measurement scenario at the Beijing University of Posts and Telecommunications.}
    \label{Scenario}
\end{figure}

\begin{figure}
    \centering
    \begin{subfigure}[b]{0.48\linewidth}
        \centering
        \includegraphics[width=\linewidth]{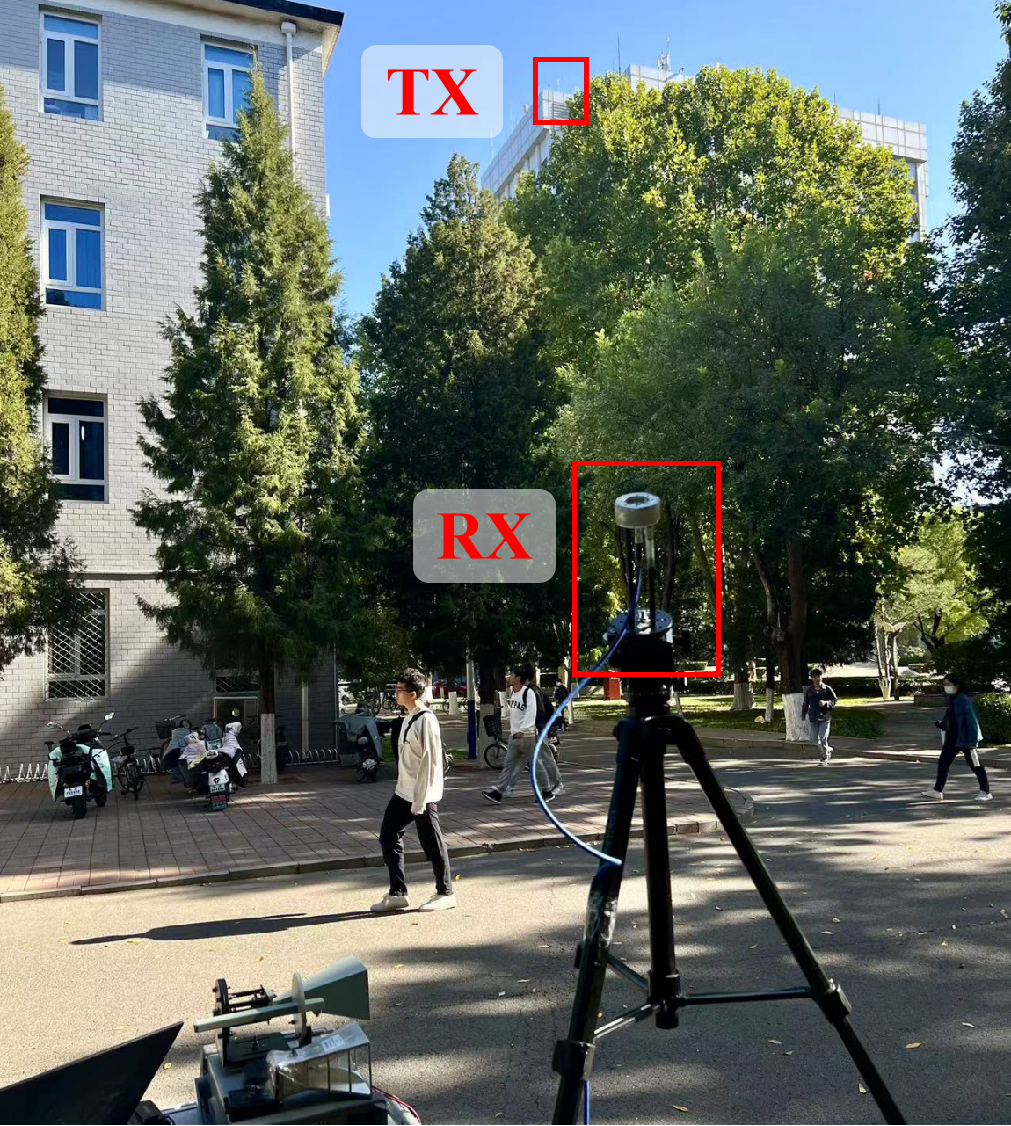}
        \caption{}
        \label{fig:image1}
    \end{subfigure}
    \begin{subfigure}[b]{0.48\linewidth}
        \centering
        \includegraphics[width=\linewidth]{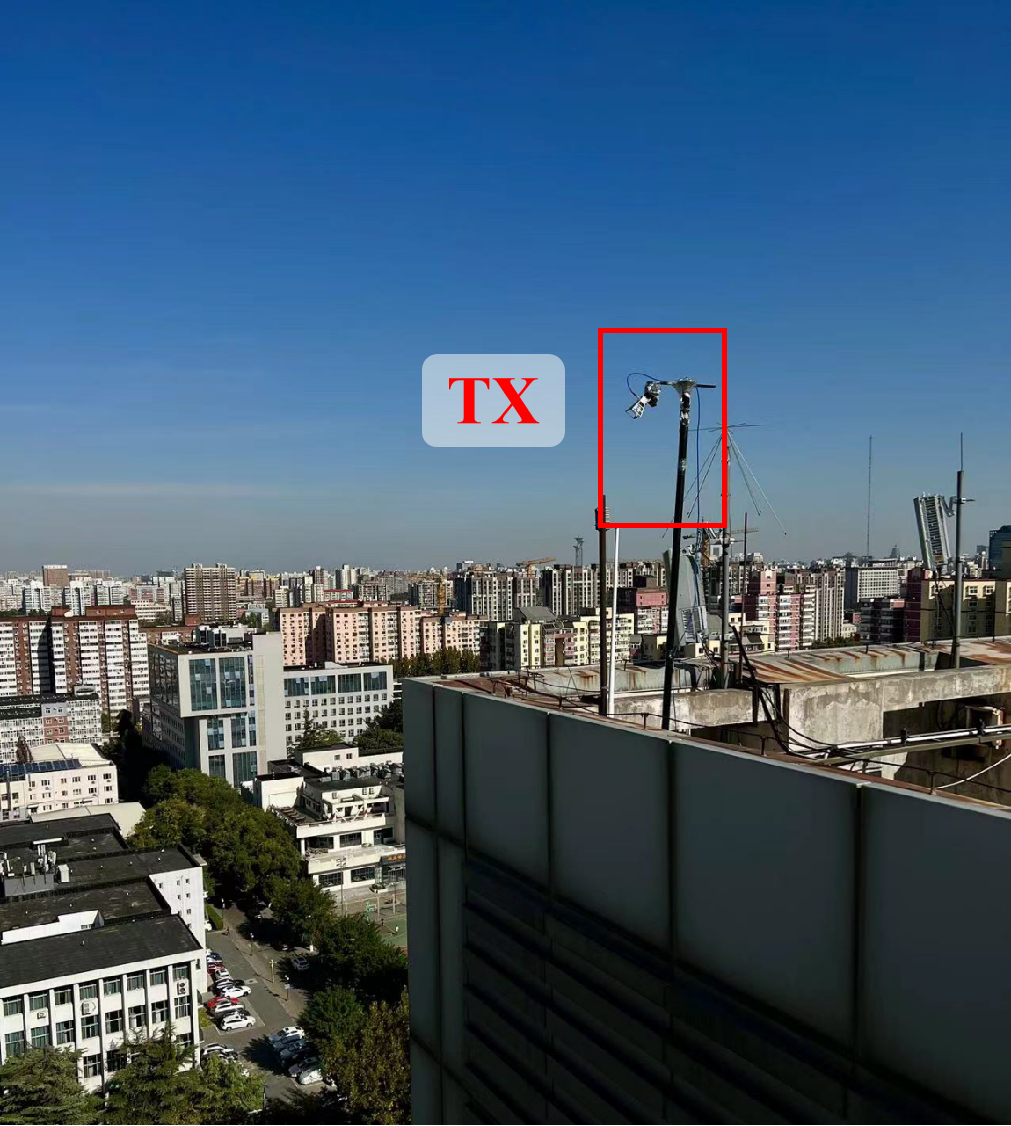}
        \caption{}
        \label{fig:image2}
    \end{subfigure}
    \caption{Photograph of Route3's environment and the positions of the TX and RX.}
    \label{Photo}
\end{figure}

The measurement campaign was conducted in a typical urban macro-cellular scenario, as depicted in Fig. \ref{Scenario}. The buildings in the scenario vary from 12 to 60 meters in height, and there is also a considerable amount of vegetation present, with vegetation heights ranging approximately from 10 to 20 meters. The measurement routes and Tx locations are also marked in Fig. \ref{Scenario}, with the 3D distance between the Tx and Rx spanning from 70.7 to 304.6 m. To prevent obstruction of the antenna beam by walls, the Tx antennas are positioned near the edges of the walls. Four transmitter antenna locations were selected based on the location of the measurement routes. Fig. \ref{Photo} shows the transmitter antenna positions for route 3. The number of measurement points, transmitter antenna positions, and 3D distance range for each measurement route are listed in Table \ref{Routeparameter}.

At each measurement point, a total of 7 frequencies were measured: 10.2, 11.2, 11.8, 12.4, 13, 13.6, and 14.8 GHz in FR3 band. For each frequency, a total of 50 channel snapshots were captured to be averaged to reduce noise. The detailed configuration of the channel sounder is outlined in Table \ref{parameter}. Given the large number of measurement frequencies, a frequency sweep is performed around all the measured frequencies before the measurement to ensure precision. This step helps detect any potential interference signals. Our analysis revealed no signals that could affect the measurements.

\begin{table}
\renewcommand\arraystretch{1.5}
\caption{Measurement route configuration.}
\begin{center}
\begin{tabular}{m{1cm}<{\centering}m{2cm}<{\centering}m{1.5cm}<{\centering} m{1.5cm}<{\centering}}
\toprule 
\textbf{Route} & \textbf{Distance range}&\textbf{Number of points} &\textbf{TX position} \\
\hline
1 & 70.7 - 213.3 m  & 34 & TX 1\\
2 & 154.5 - 249.7 m & 24 & TX 1 \\
3 & 103.2 - 304.6 m& 31 & TX 2 \\
4 & 269.3 - 193.6 m & 26 & TX 2\\
5 & 71.6 - 104.5 m& 27 & TX 3\\
6 & 82.2 - 123.9 m & 21 & TX 4\\
\bottomrule
\end{tabular}
\label{tab1}
\end{center}
\label{Routeparameter}
\end{table}

\begin{table}
\renewcommand\arraystretch{1.5}
\caption{Measurement system parameters.}
\begin{center}
\begin{tabular}{m{4cm}<{\centering} m{4cm}<{\centering}}
\toprule 
\textbf{Parameter} & \textbf{Value} \\
\hline
TX height & 62.5 m \\
RX height & 1.85 m \\
Centre frequency & 10.2, 11.2, 11.8, 12.4, 13, 13.6, 14.8 GHz \\
Symbol rate & 200 Msymbol/s \\
Bandwidth & 400 MHz \\
TX antenna & Dual-ridge horn antenna\\
RX antenna & Biconical antenna\\

\bottomrule
\end{tabular}
\label{tab1}
\end{center}
\label{parameter}
\end{table}

\subsection{Characteristic Analysis}
\begin{figure}[htb]
    \centering
    \includegraphics[width=0.45\textwidth]{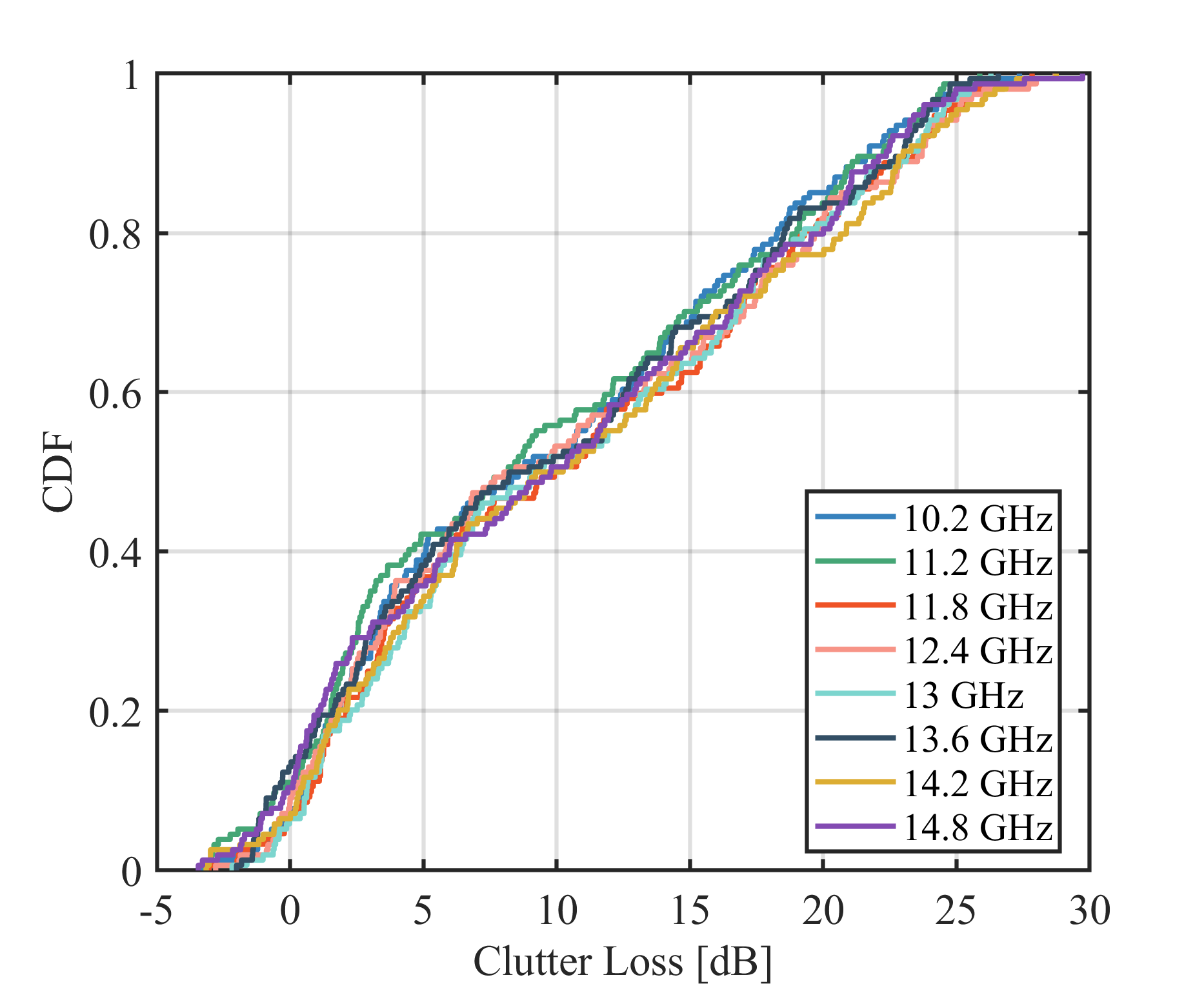}
    \caption{The clutter loss at 10.2-14.8 GHz in FR3 band.}
    \label{clutterloss}
\end{figure}

The clutter loss, as defined by ITU,  can be calculated by \cite{2108}:

\begin{equation}
    L_{clu} = L_{mea} - 32.44 - 20 \log _{10} f_c-20 \log _{10} d_{3-D} \mathrm{,}
\end{equation}
where $L_{clu}$ is the clutter loss in dB, $L_{mea}$ is the measured path loss in dB, $f_c$ is the carrier frequency in GHz, and $d_{3-D}$ is the 3-D distance in meter between the TX and RX, respectively. 
Fig. \ref{clutterloss}presents the CDF of the clutter loss at the measured frequency points. It is evident that the clutter loss shows comparable distribution curves at different frequencies, suggesting that its dependence on frequency is minimal.

\section{PENETRATION LOSS MEASUREMENT RESULTS}
\label{sec:V}

Penetration loss is the difference in signal strength between the inside and outside of a wall at the same height above ground\cite{pn_bib6}. It is an important factor for designing wireless networks, as it impacts coverage, capacity, and service quality\cite{pn_bib7}, \cite{pn_bib8}. Since penetration loss depends on frequency and varies greatly, especially in the FR1 and FR3 bands (450 MHz to 24 GHz), studying this loss in these bands is crucial.

\subsection{  Penetration Loss of Building Materials }
\begin{figure}[!ht]
\centering
\subfloat[The LOS scene.]{
		\includegraphics[scale=0.052]{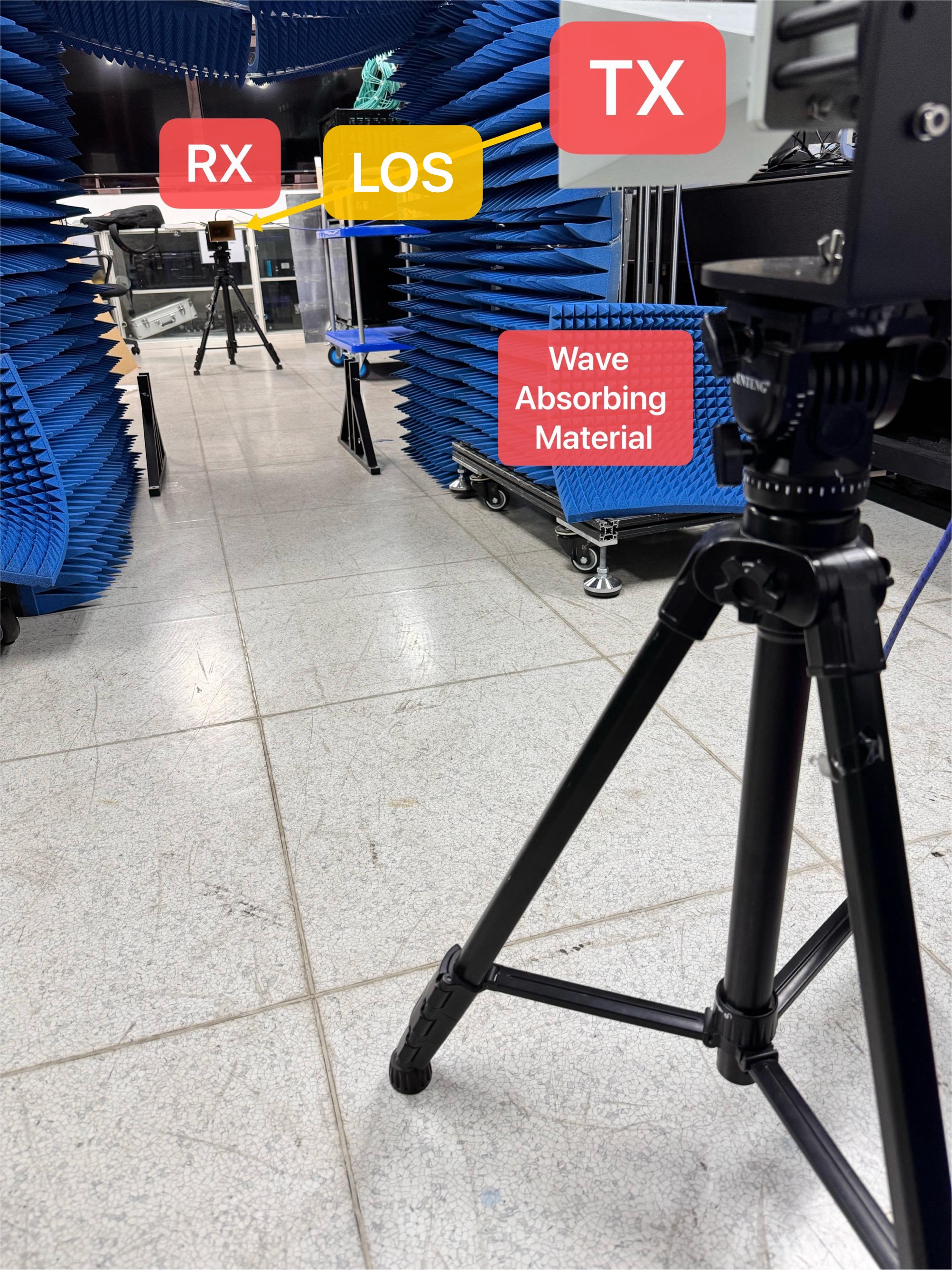}}
\hspace{0.2 cm}
\subfloat[The NLOS scene.]{
		\includegraphics[scale=0.0742]{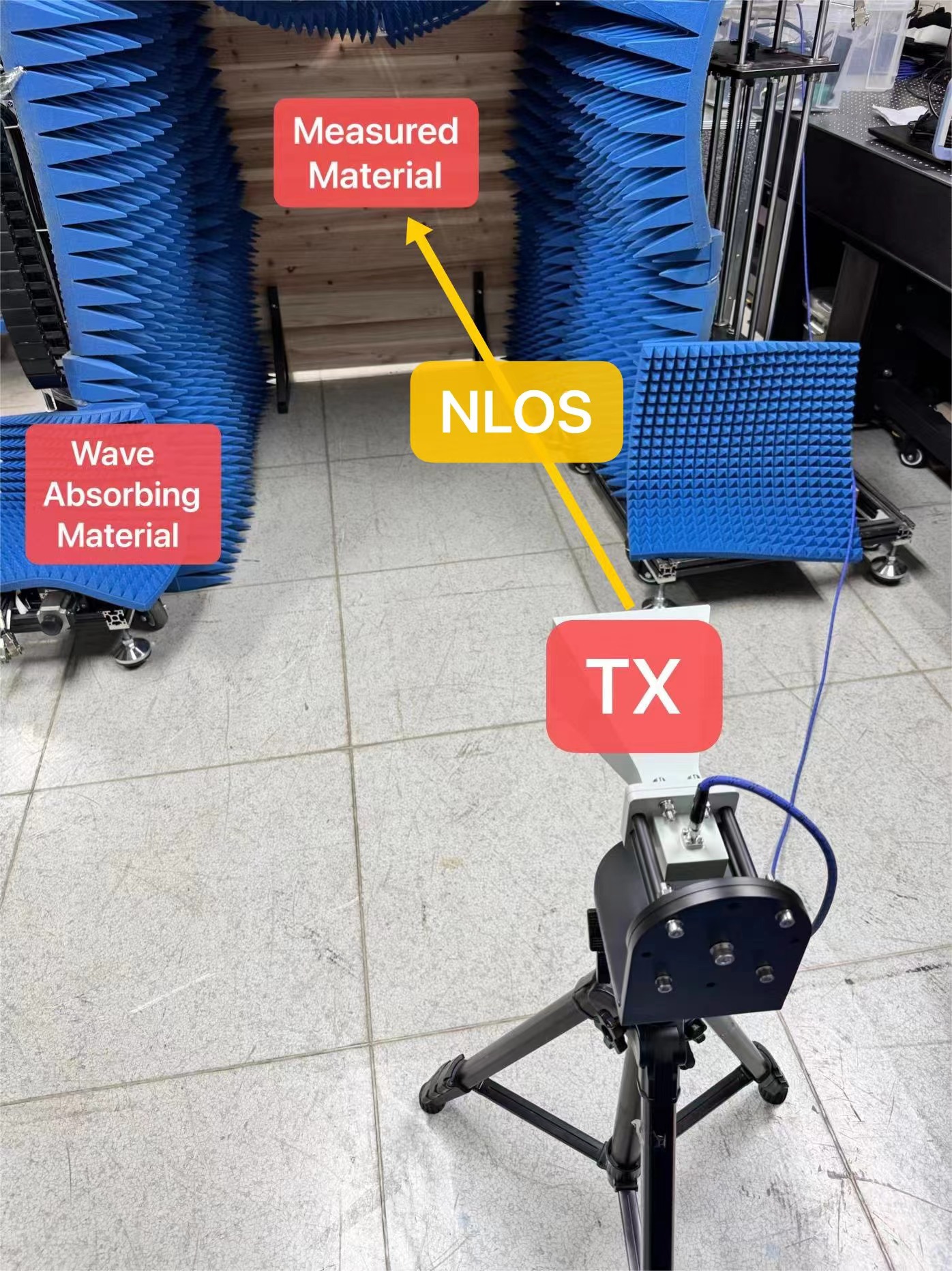}}
\caption{The penetration loss measurement scenario.}
\label{pn_fig:2}
\end{figure}

The measurement setup is as follows: Antennas for the relevant frequency bands are placed on both sides of the material and connected through a Vector Network Analyzer and cables\cite{pn_bib2}. First, the LOS path is measured without the material, as shown in Fig. \ref{pn_fig:2}(a). Then, the material is placed between the antennas, and the path loss is measured under NLOS conditions, with the signal blocked by the material, as shown in Fig.\ref{pn_fig:2}(b). The measured data are processed to calculate and visualize penetration loss, with the TR 38.901 standard included for comparison  \cite{liu2024experimentalanalysis}.

We analyze the penetration loss of materials like wood, glass, concrete, and foam boards across different frequencies. Wood’s loss increases with thickness, while glass shows varying trends, with frosted glass having lower loss. Concrete’s loss rises with frequency and differs from the TR 38.901 model. Foam boards have stable and low loss. These results highlight material differences and deviations from the standard model.

Generally, penetration loss increases with thickness, and most materials show a linear relationship with frequency. Foam causes the least attenuation, while concrete causes the most. Overall, the TR 38.901 model doesn’t predict penetration loss well in the FR1 and FR3 bands. A linear model is used to fit the measured data.
\begin{eqnarray}
PL = k*f + b,
\label{pn_formula:4}
\end{eqnarray}
where $PL$ represents the predicted penetration loss value, $f$ is the frequency, $k$ is the slope, indicating the change in $PL$ for each unit increase in f, and $b$ is the intercept. The specific details such as slope and intercept are shown in Table \ref{pn_tab:3}. For wood and foam boards, the slopes are the same for different thicknesses of the same material, with a slight difference in intercepts (0.23 dB, 0.4 dB). 

\begin{table}[t]
    \centering
    \caption{Comparison of fitting results for different materials with corresponding penetration loss models in TR 38.901.}
    \resizebox{\linewidth}{!}{%
    \begin{tabular}{|c|c|c|c|}
        \hline
        \textbf{Category} & \textbf{Name} & \textbf{Slope (k)} & \textbf{Intercept (b)} \\
        \hline
        \multirow{4}{*}{Wood} & Wooden Board 1 & 0.23 & 1.75 \\
                              & Wooden Board 2 & 0.23 & 1.52 \\
                              & Wooden Board 3 & 0.07 & 3.55 \\
                              & TR 38.901 & 0.12 & 4.85 \\
        \hline
        \multirow{3}{*}{Glass} & Double-Layer Glass & 0.30 & 2.30 \\
                                & Frosted Glass & -0.06 & 3.94 \\
                                & TR 38.901 & 0.20 & 2.00 \\
        \hline
        \multirow{3}{*}{Foam} & Foam Board 1 & -0.01 & 1.84 \\
                               & Foam Board 2 & -0.05 & 1.97 \\
                               & Foam Board 3 & -0.01 & 1.44 \\
        \hline
        \multirow{2}{*}{Concrete} & Concrete Slab & 0.95 & 9.83 \\
                                   & TR 38.901 & 4.00 & 5.00 \\
        \hline       
    \end{tabular}%
    }
    \label{pn_tab:3}
\end{table}

\begin{figure}[!ht]
    \centering
    \includegraphics[width=1\linewidth]{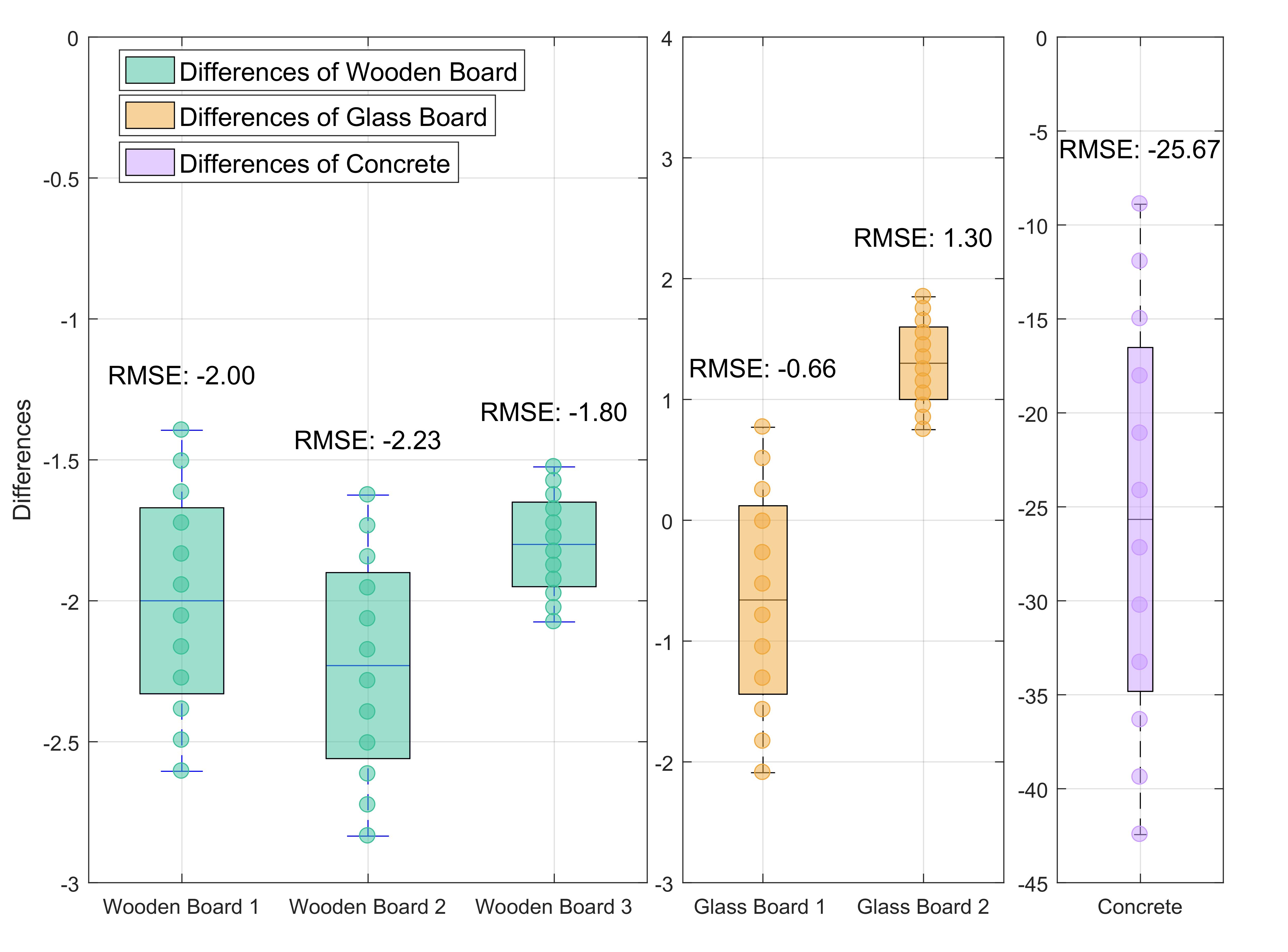}
    \caption{Difference and RMSE value distributions of penetration loss measurements between the fitted model and the standard model for different materials.}
    \label{pn_fig:4}
\end{figure}

To assess the differences between the TR 38.901 model predictions and the measurement fitting results, we calculate the differences and RMSE using the fitting results of different building materials and the corresponding TR 38.901 model. The formulas are shown in equations (\ref{pn_formula:5}) and (\ref{pn_formula:6})
\begin{eqnarray}
e_{i} = y_{i} - \hat{y_{i}}, 
\label{pn_formula:5}
\end{eqnarray}
\begin{eqnarray}
RMSE = \sqrt{\frac{\sum_{i=1}^{n} (e_{i})^{2} }{n}}, 
\label{pn_formula:6}
\end{eqnarray}
where $y_{i}$ is the fitting value of the $i$ th measured penetration loss, and $\hat{y_{i}}$ is the value of the $i$ th predicted value (the value given by the TR 38.901 model), $n$ is the total number of observations. The calculation results are shown in Fig. \ref{pn_fig:4}.

Fig. \ref{pn_fig:4} shows the differences and RMSE between the fitted and standard models for wood, glass, and concrete on a dB scale. For wood, the difference is around -2 dB, with minimal variation between boards of the same thickness. The RMSE for wood ranges from 1.81 to 2.26 dB, indicating high accuracy, with board 3 performing the best. Glass has a median difference close to 0 dB and low RMSE (1.11 to 1.35 dB). Concrete, however, shows a wider range of differences (-42.5 to -8 dB) and the highest RMSE (27.75 dB), indicating more variability.

Using difference and RMSE as metrics, we find that the standard and fitted models for wood and glass show small discrepancies. However, for concrete, the differences are much more spread out, and RMSE values are much higher. This suggests the standard model for wood is suitable in the 4–16 GHz range, but the model for concrete needs adjustment. The glass model also requires refinement as it does not fit well with a linear approach.
In conclusion, the comparison of our fitted model with the 3GPP TR 38.901 standard model shows significant differences, especially for glass and concrete materials. These differences indicate that the TR 38.901 model cannot fully capture the actual penetration loss characteristics observed in the FR1 and FR3 bands.
\subsection{Outdoor O2I Penetration}
\begin{figure}[!ht]
    \centering
    \includegraphics[width=1\linewidth]{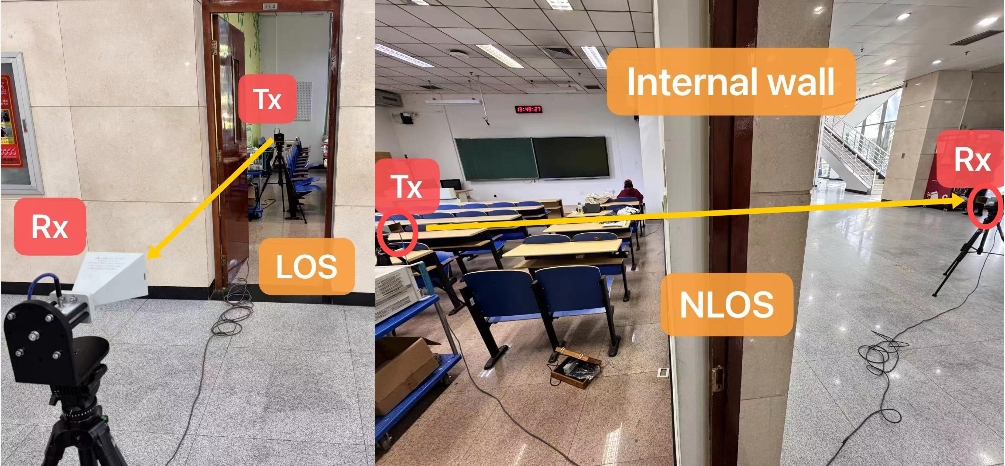}
    \caption{The penetration loss measurement scenario (Internal wall).}
    \label{pn_fig:5}
\end{figure}

\begin{figure}[!ht]
    \centering
    \includegraphics[width=1\linewidth]{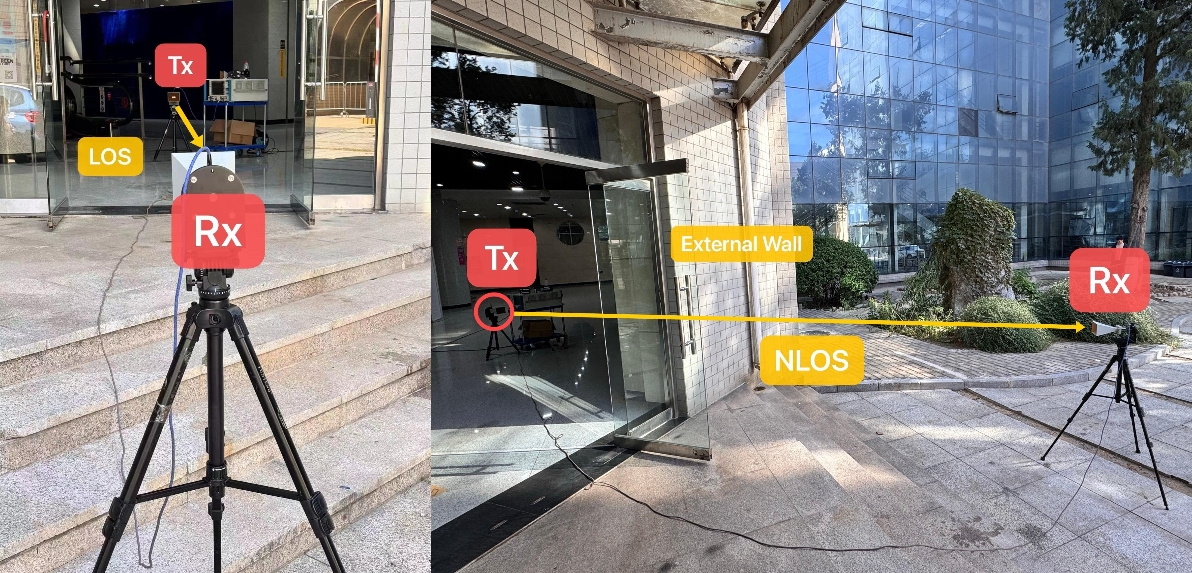}
    \caption{The penetration loss measurement scenario (External wall).}
    \label{pn_fig:6}
\end{figure}

Penetration loss is calculated as the difference between the received power levels of the unobstructed path (LOS) and the path with material obstruction (NLOS). Measurements are taken by placing antennas at both ends of the material and connecting them using a VNA. First, the LOS is measured, as shown in the left figures of Fig. \ref{pn_fig:5} and Fig. \ref{pn_fig:6}. Then, the received power is measured with different material obstacles, shown in the right figures of Fig. \ref{pn_fig:5} and Fig. \ref{pn_fig:6}. The S-parameters are averaged from ten samples for each measurement.

To reduce the impact of multipath effects, we use a bandwidth of 1 GHz, and set the center frequency to start at 4.25 GHz and increase by 1 GHz to 15.75 GHz.  For each wide-band signal, we sample 201 points and perform an inverse fast Fourier transform (IFFT) on the measured data to obtain the CIR. The path with the smallest delay is selected as the measurement result (i.e., the path directly penetrating the material under test). Finally, a curve is plotted to show the variation of penetration loss with frequency.

\begin{figure}[!ht]
\centering
\subfloat[Wooden Door.]{
		\includegraphics[scale=0.6]{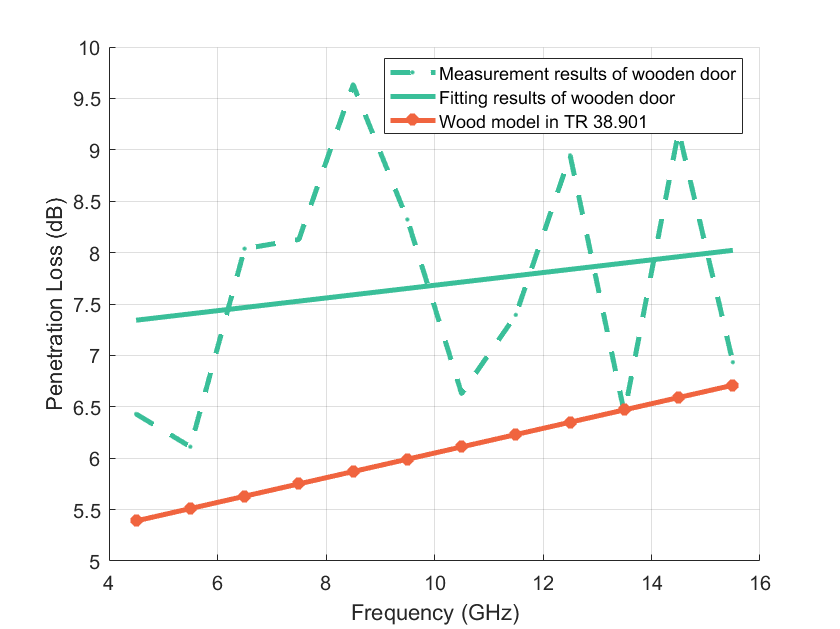}}
\\       
\subfloat[Concrete Wall.]{
		\includegraphics[scale=0.6]{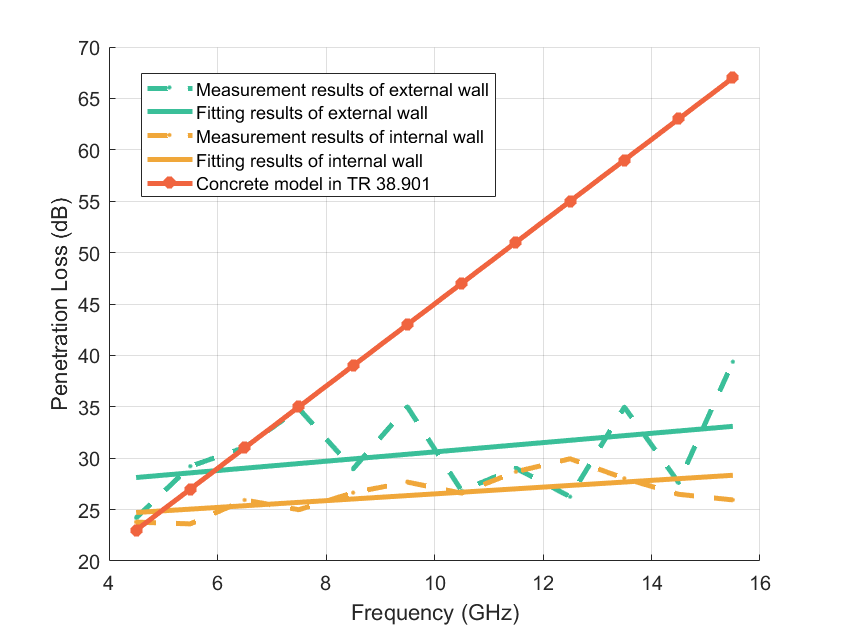}}
\caption{The penetration loss measurement scenario.}
\label{pn_fig:7}
\end{figure}
The measured data is processed to calculate the penetration loss, which is then plotted and fitted with a linear function, as shown in the figures.

The results show that the penetration loss for the materials tested generally increases with frequency and thickness. Concrete has the strongest signal obstruction compared to wood. The measured results are similar to those in TR 38.901.

For wooden doors (Fig.\ref{pn_fig:7} (a)), the fitted penetration loss is about 1.5 dB higher on average than the 3GPP standard, with slopes of 0.06 and 0.12. For walls (Fig.\ref{pn_fig:7} (b)), penetration loss varies from 0 to 11 dB. The results show that outdoor walls have about 5 dB higher penetration loss than indoor walls, likely due to the greater thickness of outdoor walls. The slope for walls also deviates from the standard, suggesting that the model may need adjustment.

\section{CONCLUSION}
\label{sec:VI}
This paper presents a comprehensive summary of the mid-band channel characteristics, within outdoor environments. Through extensive channel measurement campaigns, both far-field and near-field channel characteristics were examined, highlighting the spatial non-stationarity and frequency-dependent variations of key parameters, including path loss, delay spread, angular spread, and channel sparsity. The findings underscore the significant impact of these characteristics on the design and optimization of 6G communication systems.

Notably, our review indicates that outdoor environments exhibit pronounced spatial non-stationarity, especially in near-field scenarios, which is crucial for technologies such as XL-MIMO. Additionally, the study on ground-to-air communication revealed that clutter loss remains relatively independent of frequency, whereas penetration loss shows clear frequency dependencies, with materials such as concrete showing significant deviations from existing models like TR 38.901. These results provide essential empirical data that can guide future enhancements to channel models and the deployment of mid-band spectrum for 6G.

This survey highlights the necessity of further exploration into the characteristics of the mid-band spectrum, and the refinement of existing channel models to accommodate the unique propagation characteristics that will define future wireless communication systems.


\section*{ACKNOWLEDGMENT}
This work was supported by National Key R\&D Program of China (2023YFB2904802), Natural Science Foundation of Beijing-Xiaomi Innovation Joint Foundation (L243002), National Natural Science Foundation of China (62101069 \& 62201086), and Beijing University of Posts and Telecommunications-China Mobile Research Institute Joint Innovation Center.

\bibliographystyle{IEEEtran}
\bibliography{ref}


\end{CJK}
\end{document}